# FRACTAL MODELS OF EARTHQUAKE DYNAMICS

Pathikrit Bhattacharya[1], Bikas K. Chakrabarti[2], Kamal[1], Debashis Samanta[2]

[1]Department of Earth Sciences, Indian Institute of Technology, Roorkee-247 667, Uttarakhand, INDIA

[2]Theoretical Condensed Matter Research Division and Centre for Applied Mathematics and Computational Science, Saha Institute of Nuclear Physics, Sector-1, Block–AF, Bidhannagar, Kolkata-700 064, INDIA.

E-mail addresses: pathipes@iitr.ernet.in (Pathikrit Bhattacharya), bikask.chakrabarti@saha.ac.in (Bikas. K. Chakrabarti), kamalfes@iitr.ernet.in (Kamal), debashis.samanta@saha.ac.in (Debashis Samanta)

Our understanding of earthquakes is based on the theory of plate tectonics. Earthquake dynamics is the study of the interactions of plates (solid disjoint parts of the lithosphere) which produce seismic activity. Over the last about fifty years many models have come up which try to simulate seismic activity by mimicking plate plate interactions. The validity of a given model is subject to the compliance of the synthetic seismic activity it produces to the well known empirical laws which describe the statistical features of observed seismic activity. Here we present a review of two such models of earthquake dynamics with main focus on a relatively new model namely 'The Two Fractal Overlap Model'.

# 1 Introduction

### 1.1 Earthquake statistics

The overall frequency distribution of earthquakes is given by the Gutenberg-Richter (GR) Law [1] which states

$$log\ N(m) = a - bm, \qquad (1.1)$$

where $N(m)$ is the frequency of earthquakes with magnitude greater than $m$ occurring in a specified area. The constant $b$, the so called '$b$-value', has some regional variation (the value of the exponent $b$ has been seen to change from one geographical region to another) but the universally accepted value of $b$ is close to unity. The constant '$a$' is a measure of the regional level of seismicity. However owing to the log-linear relationship between seismic energy released and the magnitude of the earthquake, there is another form in which the Gutenberg-Richter law is stated:

$$N(\varepsilon) \sim \varepsilon^{-\alpha} \qquad (1.2)$$

where $N(\varepsilon)$ is defined in analogy to the previous form but for events which release energy greater than $\varepsilon$. This is due to the fact that usually magnitude is defined as logarithm of the trace amplitude on a seismogram and hence bears a log-linear relationship with energy. The temporal distribution of aftershocks of magnitude $m$ greater than or equal to some threshold value $M$ is given empirically by another well known power law, namely the Omori Law [2], saying

$$\frac{dN(t)}{dt} = \frac{1}{t^p}, m \geq M \;. \qquad (1.3)$$

Here $dN(t)/dt$ gives the rate of occurrence of aftershocks at time $t$ after the occurrence of the mainshock. The value of the exponent $p$ is close to unity for tectonically active regions though a large range of variation in the $p$ value has been observed [3].

## 1.2 Modeling earthquake dynamics

The principal objective in constructing models of earthquake dynamics is to reproduce the above two empirical (statistical) laws by simulating the dynamics of a fault or of a system of interconnected faults. Different types of models have been proposed to capture this dynamics which focus on different aspects of fault dynamics. One class mimics the dynamics by slowly driving an assembly of locally connected spring-blocks over a rough surface. This essentially captures the stick slip scenario involved in generation of earthquakes. The first successful model of this kind was proposed by Burridge and Knopoff [4]. This model and all its variants [5, 6] have been reliably shown (numerically) to recreate the GR Law but the Omori Law has not been clearly demonstrated from this class of models. The underlying principle for this class of models has been found to be Self Organized Criticality [7]. There is another traditional class of models based on the mechanical properties of deformable materials that break under a critical stress. Fibre bundle models discussed in Section 4 of this review are typical of this class.

The main class of models that we will discuss here are a relatively new type. This class of models deals with the fractal geometry of fault surfaces. We shall discuss in the next section some of the available observations indicating that fault surfaces are fractals and how faults are distributed in a fault zone with a fractal size-distribution. These are two very well established facts. Naturally, a few of the geometrical models of earthquakes capture the fractal effects of one fault surface sliding over the other by considering two fractals sliding over each other and by taking into account the stresses developed and released due to such overlaps. Fig. 1.1 shows a cartoon depicting this scenario. This is the basic motivation behind fractal overlap models. There have been attempts at using random fractional Brownian profiles as the fractals involved (in the so called Self-affine Asperity Model) in [8, 9]. The model yields the GR law readily and relates the '*b-value*' to the geometry of the fault. A more generalized version of the model discussed in [9] also recreates the Omori law but with a universal exponent. But in nature the exponent value varies considerably. Also, the exponent is very different in value from the exponent observed for real earthquakes (for this Self-affine Asperity Model [9] the value of the exponent is 0.37, while in nature we observe values close to unity for seismically active zones as mentioned before). Our focus though, will be on yet another geometric model which has been reasonably successful in capturing most of the observed statistical features of earthquake processes reproducing values of the parameters of these empirical laws much in agreement to what we see in nature. We call this the 'Two fractal overlap model'. The simplest scenario of a fractal sliding over its complementary set involves a Cantor set sliding over its complement. But the scenario considered here is even more simplified. We consider the overlap statistics of a Cantor set sliding over its replica. Although the model does not, to start with, consider a real fault profile the main strength of the model lies in the fact that it is completely analytically tractable and gives all the well established statistics that real earthquakes demonstrate. We will, through the length of this discussion, show these results and compare them with real earthquake data. The reader will readily recognize that these results require a knowledge of no more than high school mathematics to derive and in simplicity lies the true strength of this model.

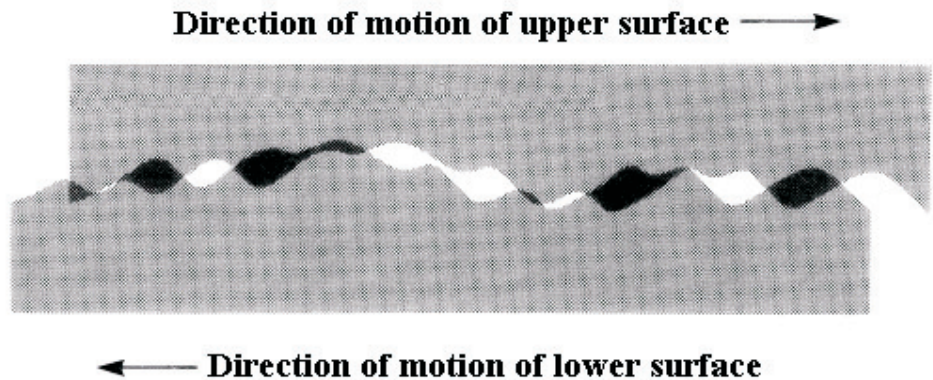

*Fig. 1.1 A cartoon showing overlap of two fractal surfaces. The sticking is due to interlocking of the asperities. Stress energy is accumulated and released at every slip (Adapted from* [10]*).*

1.3 Fractal faults

**A) Fractal geometry of fault surfaces**

Before we undertake a study and modeling of seismic activity, it is of interest for the general readers to know the uses of the terms like fractures, joints and faults. Any crack or fissure on the surface of a rock is a fracture. If the two blocks separated by the fracture are laterally displaced creating a plane across which the rock beds are discontinuous then, in strict terms of structural geology, the locus of the discontinuity in the various rock beds is the fault. Fig. 1.2 shows a fault exposure in the Dixie Valley in the United States. If there has been no lateral offset across the fracture then the structure is generally referred to a joint. Faults and joints often do not come singly but in a complex system of interconnected structures. Such a system of interconnected faults is called a fault zone. In other words, it is basically a highly fractured system of fault networks all of which have been formed by the same tectonic process. It has for long been suggested that fractured rock surfaces are fractals. The fractal geometry implies a balance between two competing processes: strain weakening and strain hardening. This balance is critically tuned to produce neither positive nor negative feedback mechanisms during deformation. In such a case, the long-term deformation is accommodated statistically, at all time intervals, by structures that have no preferred size scale, i.e., structures following a scale free (due to the lack of feedback) frequency-size distribution. Fractal geometry has been reported to characterize brittle deformation structures in the crust over several bands of length scales, from regional fault networks through main traces of individual faults to the internal structure of fault zones.

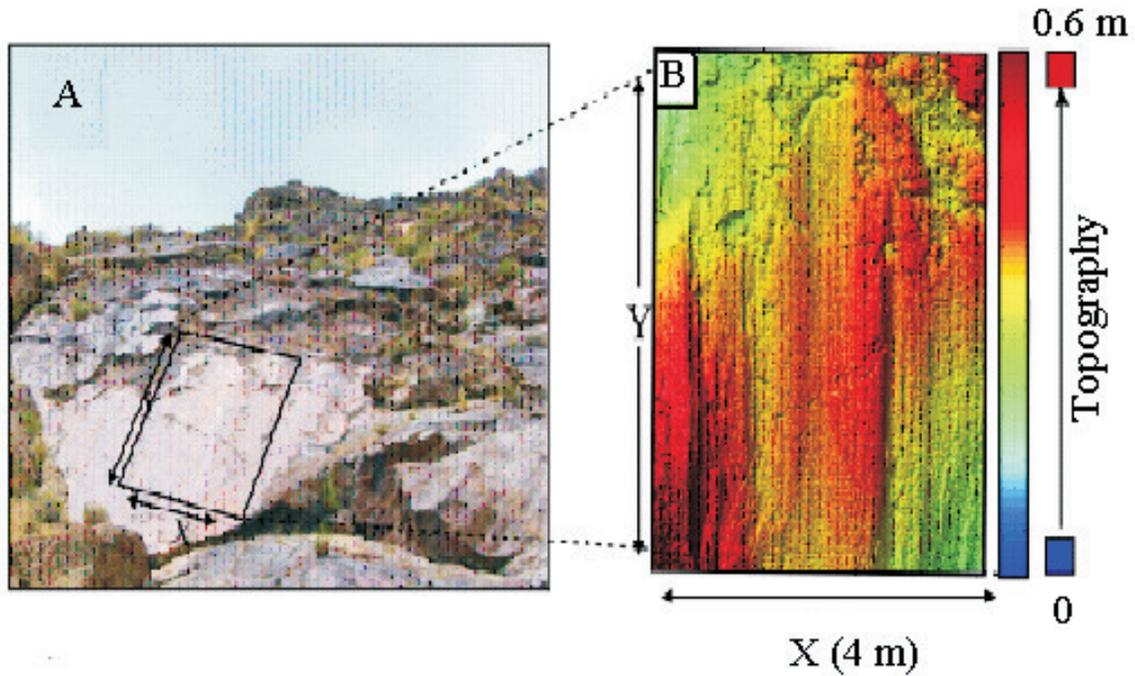

*Fig. 1.2 A) Section of a partly eroded slip surface at the Mirrors locality on the Dixie Valley fault. B) LiDAR fault surface topography as a color-scale map rotated so that the X-Y plane is the best-fit plane to the surface (Adapted from [11]).*

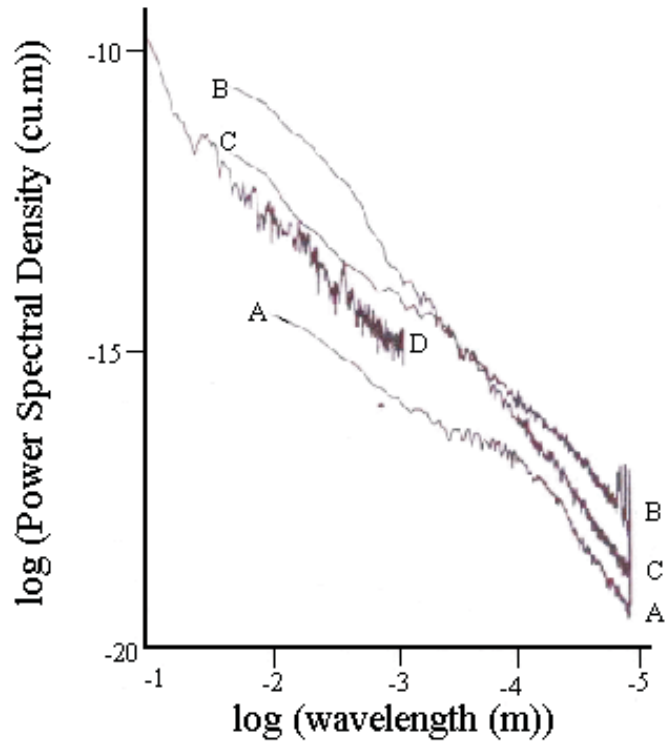

*Fig. 1.3 Power spectra for the fault surfaces studied in the Dixie Valley by* [13]. *A, B, C are from 10 – 20 mm long lab profiles. D is part of a spectrum from a 1 m long field profile. A – smoothest, unweathered hand sample of surface. B – sugary weathered surface. C – surface that apparently a is composite of sub-parallel surfaces (Adapted from* [13]*).*

In fact, fault surfaces are fractals. It was shown by Brown and Scholtz [12] that the surfaces of joints are fractal. They studied the surface topography of naturally occurring joints by analyzing the power spectra of the profiles. They studied fresh joints (a fresh surface in structural geological context implies an unweathered surface) in both sedimentary and crystalline rocks, a frictional wear surface due to glacial activity and a bedding plane surface. The power spectrum of all these surfaces showed a 'red noise' spectrum over the entire spatial frequency bandwidth employed in the study with the amplitude falling off 2 to 3 orders of magnitude per decade increase in spatial frequency.

This was explained using a fractal model of the topography. The dimension *D* was found to vary with spatial frequency. Power et al. [13] did a similar analysis on the surface of faults in the western United States and found fault surfaces to be fractal over eleven orders of magnitude in wavelength. They found that the amplitude of the spectrum increased roughly in proportion to the wavelength under consideration. The power spectra for the fault surfaces in Dixie Valley (Western United States) are shown in Fig. 1.3 as reported in [13]. Such studies have been strengthened by modern techniques of imaging like the LiDAR profile shown in Fig. 1.2. The topography of fault surfaces is now generally considered as fractal. So it is very reasonable to consider the movement of fault surfaces on and relative to one another as two fractals sliding over one another. This forms the basis of our motivation behind studying the overlap statistics of a Cantor set sliding over its replica.

**B) Frequency-size distribution of faults**
At the smallest scale, it has been shown in [14] that the frequency-size distribution of microfractures developed under stress in an unfractured and stress-free rock (granite in this case) is a power law. The GR law itself is a power law. In fact in [15] it has been shown that the GR law is exactly equivalent to a fractal distribution of seismic activity versus rupture size (rupture size is the area of the rupture for the event). The fractal dimension of this distribution $d_f$ and the '*b-value*' are related as $d_f = 2b$ [16]. A reasonable hypothesis now accepted by almost all seismologists is that each fault has a characteristic earthquake and a fractal distribution of earthquakes implies a fractal distribution of faults. This has in fact been shown by many authors by analyzing the spatial distribution of various fault networks.

The frequency-size distribution of faults belonging to a given fault network is usually implemented by using either Richardson plots or by the (more prevalent now-a-days) box-counting technique. Aki et al. [17] and Scholz et al. [18] independently studied the fractal geometry of various segments of the San Andreas Fault system. Hirata [19] did the same for fault systems in Japan; Villemin et al. [20] have studied the frequency-size distribution of faults in the Lorraine coal basin in France; Idziak and Teper [21] have done similar work for fault networks in the upper Silesian coal basin, Poland; Angulo-Brown et al. [22] studied the distribution of faults, fractures and lineaments over a region on the western coast of Guemero, Mexico. Sukmono et al. [23, 24] have studied the fractal geometry of the Sumatra fault system. Sunmonu and Dimri [25] studied the fractal geometry and seismicity of Koyna-Warna, India; Roy and Ram [26] have studied the fractal geometry of fault networks in the Kutch region in India; Nanjo et al. [27] have studied a system of active faults in the aftershock region of the Fukui earthquake (28/6/1948, Epicenter latitude: 36.2, Epicenter longitude: 136.2, M = 7.1. Here M is the reported magnitude of the event. M mentioned in this format will from now on imply reported magnitude of the event unless otherwise specified.). Fig. 1.4 shows the tectonic map of the Kutch region studied in [26]. Fig. 1.5 (A) shows the fault networks in the Kutch region in details, again used in [26] for box counting purposes. Fig. 1.5 (B) shows the frequency-size distribution obtained by box-counting in the various blocks marked in Fig. 1.5 (A). To interpret the plot, we may look into the method of box-counting (see for example [26]). In this method the fault on the map was initially superimposed on a square grid size $r_0$. The unit square of area $r_0^2$ was sequentially divided into small squares of linear size $r_1 = r_0/2, r_2 = r_0/4, r_3 = r_0/8....$. The number of squares or boxes $N(r_i)$ of linear size $r_i$ intersected by at least one fault line are counted each time. If the fault system has a self-similar structure, then

$$N(r_i) = r_i^{-d_f} \tag{1.4}$$

where $d_f$ is interpreted as the fractal dimension of the fault system. The fractal dimension $d_f$ was determined from the slope of the $\log N(r_i)$ versus $\log(1/r_i)$ plot. A detailed discussion on the various types of these studies by different groups may be found in [16].

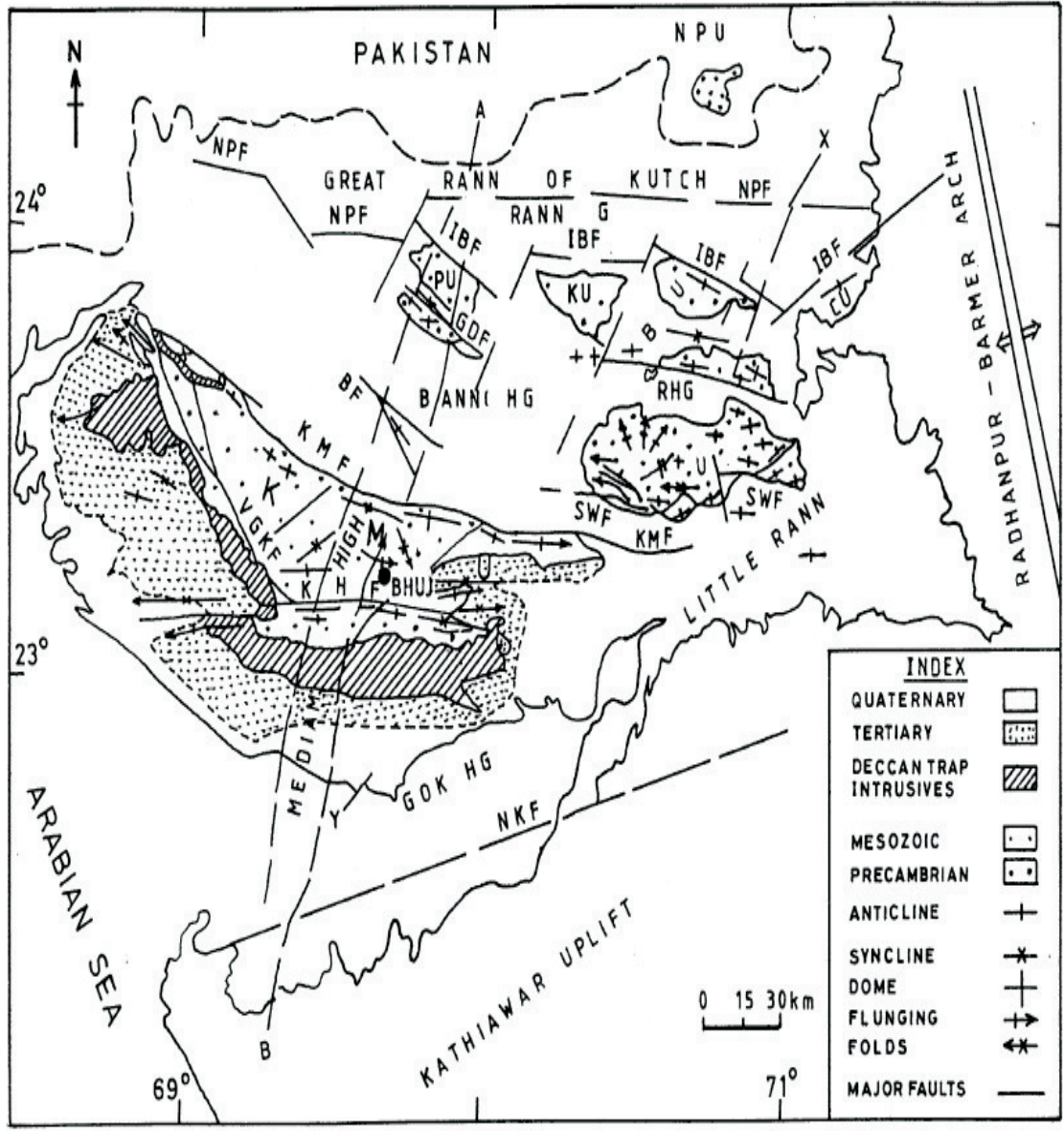

*Fig. 1.4 Structural map of the Kutch region in India (Adapted from Roy et al [26].) showing the major faults of the region.*

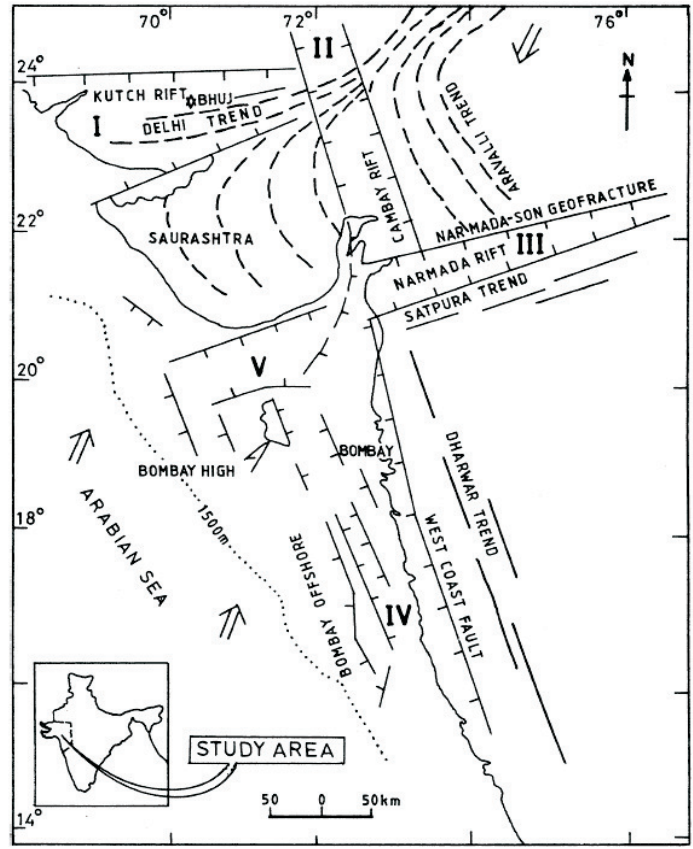

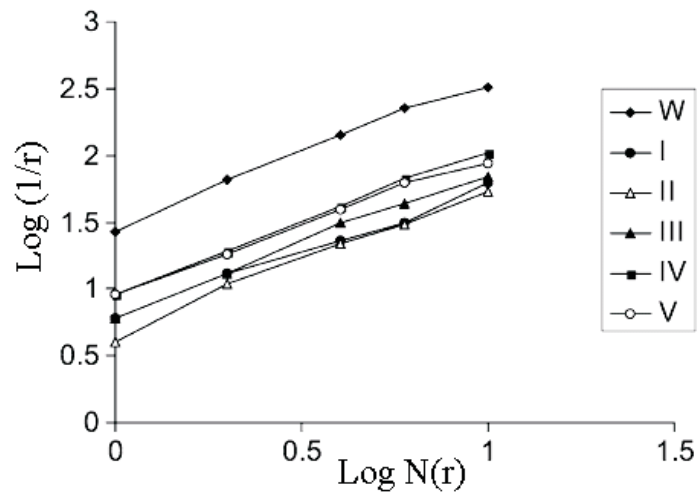

*Fig. 1.5 A) The tectonic map of the Kutch region studied in [26]. The various blocks are shown in the map. B) The frequency – size distribution for the fault systems for the different blocks marked in the map (Taken from [26]). W stands for the whole region.*

## 2 Two Fractal Overlap model

### 2.1 The model

As discussed already in Section 1, earthquakes are physically caused by the slip movements of adjacent fault planes along the contact of hanging wall and footwall asperities and the release of the stress energy accumulated due to friction during the period of sticking. But as (i) faults surfaces are fractals (ii) friction is purely a surface phenomenon and (iii) the motion is, in general, in a given direction, the process that causes release of the stored elastic energy can be analyzed effectively in one dimension. Therefore a fractal embedded in one dimension can provide us a suitable geometry to investigate the overall process. The sliding of one fractal over another thus would mimic a stick-slip scenario where the slip occurring after a stick would effectively be the physical process through which the stored energy would be released. The one dimensionality of the problem previously discussed means that we have to consider a fractal embedded in 1-D and the natural choice is a Cantor set. This is especially valid due to the fact that the projection of any fractal surface in a 1-D space is clearly a Cantor set (albeit a random one in most cases we encounter in nature). But for the sake of analytical tractability of the process we adopt, in this case, the middle third removal algorithm to generate it (fractal dimension is $\log 2/\log 3$). The dynamical model involves one such Cantor set moving with uniform relative velocity over its replica and one looks for the time variations of the measure of the overlapping sets common between the two at any instant of time. The model was initially given by Chakrabarti and Stinchcombe [28]. Chakrabarti and Stinchcombe tackled the problem of determining the overlap statistics using a renormalization group method, which has been discussed in Appendix A. We present next a modified analysis, following Bhattacharya [29]. The model considered here, as we said earlier, employs two Cantor sets of the same generation and dimension sliding over each other with uniform velocity as shown in Fig. 2.1. For the *n*-th generation, the step size is $1/3^n$ and the time taken to cover each step is taken as unity. Stress energy is accumulated at each overlap of the non-empty intervals of the upper (moving) Cantor set with the non-empty intervals of the lower (stationery) Cantor set. The extent of such overlaps (the number of such overlapping non empty intervals) is represented by the 'overlap magnitude'. This measure may represent the stress (or stress energy) accumulated due to friction within the surfaces which gets released through slips. The energy released at each such 'slip' is proportional to the overlap magnitude during the 'stick' period. We therefore need to evaluate the overlap time series. At any finite generation, the time series is exactly solvable in this model.

## 2.2 Analysis of the time series

As mentioned already, we present here a modified version of the analysis of the Chakrabarti Stinchcombe model by Bhattacharya [29]. We employ periodic boundary conditions to formulate the time series. The overlap magnitude is evaluated in terms of the number of pairs of non-empty intervals overlapping at a time. Therefore the overlap magnitude $O_n(t)$ can only assume values in a geometric progression given by $O_n(t) = 2^{n-k}$, $k = 0, 1, \ldots, n$. Clearly $O_n(0) = 2^n$ and, due to the periodic boundary conditions, taking unit time to be the time required to take one step of size $1/3^n$ we obtain

$$O_n(t) = O_n(3^n - t), \; 0 \leq t \leq 3^n \tag{2.1}$$

owing to the symmetric structure of the finite generation Cantor set.

A detailed analysis of the time series reveals a straightforward recursive structure. If we simulate the overlap time series for the $n$-th generation, after $3^{n-1}$ time steps we have the overlap time series for the $(n-1)$-th generation. Again after $3^{n-2}$ time steps beginning from the $3^{n-1}$ time steps previously taken we have the overlap time series for the $(n-2)$-th generation and recursively so on. In other words the entire time series for the 1st generation ($n = 1$) is contained in the time series for the 2nd generation ($n = 2$) starting from the time step $t = 3$ (of the 2nd generation time series) and ending at the time step $t = 6$ (of the 2nd generation time series). Again the entire 2nd generation time series is contained in the 3rd generation time series starting from the time step $t = 9$ (of the 3rd generation time series) and ending at the time step $t = 18$ (of the 3rd generation time series). Also the entire 1st ($n = 1$) generation time series is contained in the 3rd generation ($n = 3$) time series starting from the time step $t = 12$ (of the 3rd generation time series)

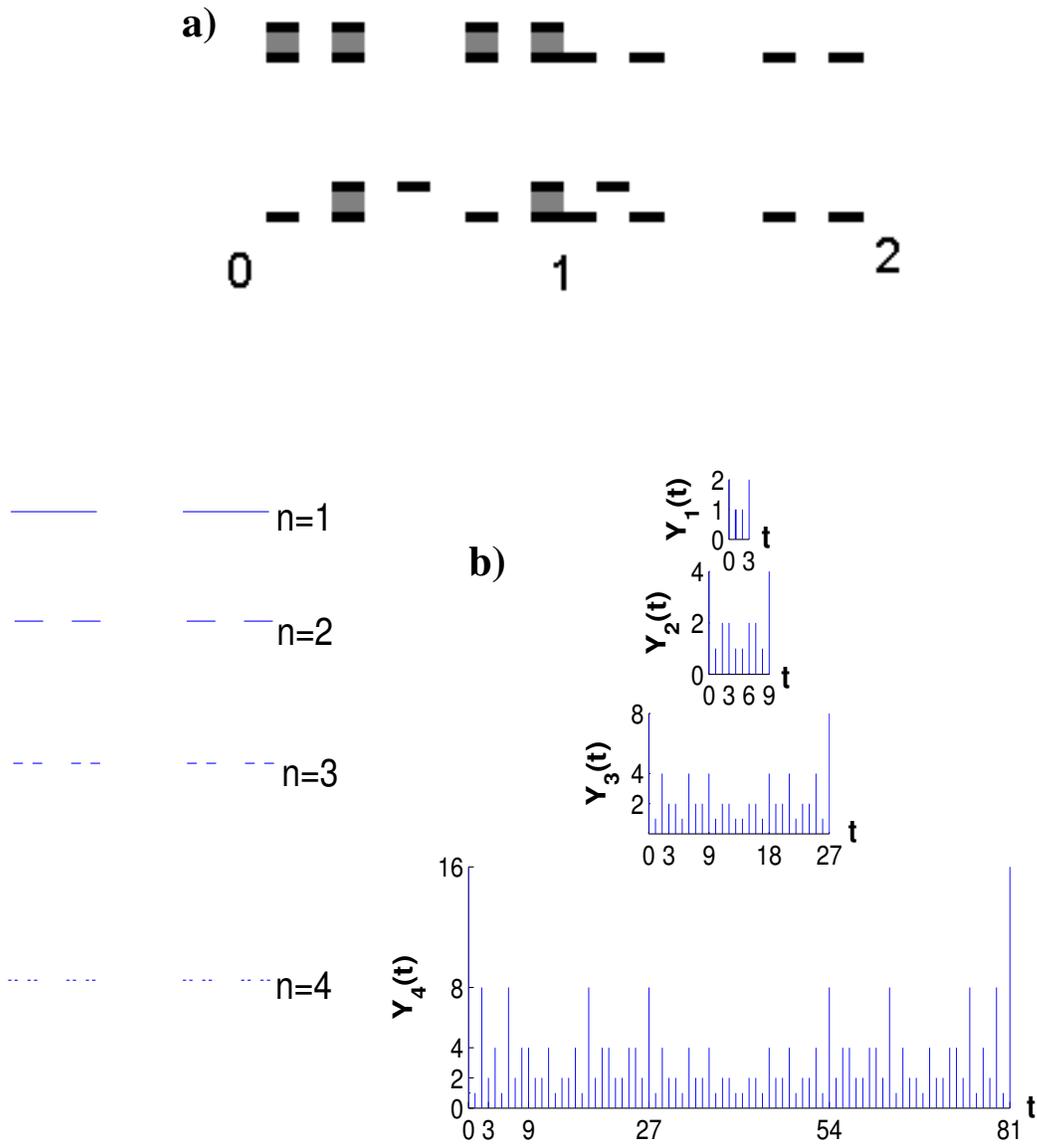

*Fig. 2.1 a) A realization of the model for the second generation at t=0 and at t=2. The overlapping segments are shaded in grey. The lower Cantor set is repeated between 1 and 2 to employ the periodic boundary condition. The upper Cantor set slides over the lower. b) The recursive structure of the time series for overlap $Y_n(t)$ or $O_n(t)$ the first four generations in the fractal-fractal overlap model. The respective Cantor set generations are shown on the left. It is noticeable that the time series of all preceding generations are embedded within the time series at a given generation.*

and ending at the time step $t = 15$ (of the $3^{rd}$ generation time series). This nested recursive structure is present throughout the time series of any $n$-th generation. Generalizing, we may state that the entire time series of the (*n*-1)-th generation is contained in the time series of the *n*-th generation starting from the time step $t = 3^{n-1}$ (of the *n*-th generation time series) and ending at the time step $t = 2 \times 3^{n-1}$ (of the *n*-th generation time series). Again, the entire time series of the (*n*-2)-th generation is contained in the time series of the *n*-th generation starting from the time step $t = 4 \times 3^{n-2}$ (of the *n*-th generation time series) and ending at the time step $t = 5 \times 3^{n-2}$ (of the *n*-th generation time series). Again, the entire time series of the (*n*-3)-th generation is contained in the time series of the *n*-th generation starting from the time step $t = 13 \times 3^{n-3}$ (of the *n*-th generation time series) and ending at the time step $t = 14 \times 3^{n-3}$ (of the *n*-th generation time series) and so on. This can be understood very clearly from the illustrations in Fig. 2.1. The details of the derivation are given in Appendix B.

There is however a finer recursive structure in the time series that leads to the analytical evolution of the number density distribution. At any given generation *n*, a pair of nearest line segments form a doublet and there are $2^{n-1}$ such doublets in the Cantor set. Within a given doublet, each segment is two time steps away from the other segment. This means that an overlap of $2^{n-1}$ occurs when one of the sets is moved two time steps relative to the other. Similarly, an overlap of magnitude $2^{n-1}$ also occurs if one considers a quartet and a relative shift of $2 \times 3$ time steps between the two Cantor sets. Again we can consider an octet and a relative shift of $2 \times 3^2$ time steps to obtain an overlap of magnitude $2^{n-1}$. In general if we consider pairs of blocks of $2^{r_1}$ nearest segments ($r_1 \leq n-1$), an overlap magnitude of $2^{n-1}$ occurs for a relative time shift of $2 \times 3^{r_1}$ time steps:

$$O_n(t = 2 \times 3^{r_1}) = 2^{n-1}; \quad r_1 = 0,......,n-1. \quad (2.2)$$

The complementary sequence is obtained using (2.1). We can create such rules for each of the possible overlap magnitude values $O_n(t) = 2^{n-k}$. Rules like these give us the frequency distribution of overlap magnitudes. For example, from (2.2) we can see that as $r_1$ can have *n* possible values. Also, for each of these times at which an overlap of magnitude $2^{n-1}$ occurs we have another time step in the complementary sequence (due to

(2.1)) at which again an overlap of magnitude $2^{n-1}$ occurs. Therefore the frequency of occurrence $N(O_n)$ of an overlap magnitude $O_n = 2^{n-1}$ is $2n$, that is $N(O_n = 2^{n-1}) = 2n$. The complete distribution can be obtained by studying the aforementioned recursive structure carefully (the mathematical details are given in Appendix B) and using simple combinatorics. The probability distribution of overlap magnitudes for the model comes out to be a binomial distribution:

$$Pr(2^{n-k}) = \binom{n}{n-k}\left(\frac{1}{3}\right)^{n-k}\left(\frac{2}{3}\right)^{k} \quad (2.3)$$

where $Pr(O_n) = \dfrac{N(O_n)}{3^n}$, that is $Pr(O_n)$ gives the probability of occurrence of an overlap of magnitude $O_n$ in a total of $3^n$ time steps. Now, remembering that the overlap magnitude $2^{n-k}$ is proportional to energy we can put $\log_2 Y_n = n - k = m$ where $m$ is the magnitude analog for the model. It must however be kept in mind that while analyzing the model $n$ is a constant as we are considering the model at a specific generation number and $m$ changes as $k$ changes. Then the frequency distribution for the model in terms of magnitude becomes

$$Pr(m) = \binom{n}{m}\left(\frac{1}{3}\right)^{m}\left(\frac{2}{3}\right)^{n-m} \quad (2.4)$$

## 2.3 The Gutenberg Richter law

In the limit of large $n$ the Cantor set becomes a true mathematical fractal and we have the standard normal approximation of (2.4) which gives (see Appendix B for a more detailed explanation):

$$F(m) = \frac{3}{2\sqrt{n\pi}} \exp\left[-\frac{9}{4}\frac{(m-n/3)^2}{n}\right] \quad (2.5)$$

where $F(m)$ now gives the probability density function for magnitude. Now to obtain the GR law analog from this distribution we have to integrate $F(m)$ from $m$ to $\infty$ to obtain the cumulative distribution function $F_{cum}(m)$. Neglecting terms with coefficients of the order

of $1/n\sqrt{n}$ and higher we obtain the cumulative distribution function for magnitude $m$ and above as

$$F_{cum}(m) = \frac{3}{2\sqrt{n\pi}} \exp(-n/4) \exp\left[-\frac{9m^2}{4n} + \frac{3m}{2}\right](m - n/3). \qquad (2.6)$$

Now, in the large magnitude limit, as the magnitude $m$ in the model cannot exceed $n$, the term $m^2/n \sim m$ and hence effectively (2.6) becomes

$$F_{cum}(m) = \frac{3}{2\sqrt{n\pi}} \exp(-n/4) \exp\left[-\frac{3m}{4}\right](m - n/3). \qquad (2.7)$$

On taking log of both sides of (2.7) we obtain

$$\log F_{cum}(m) = A - \frac{3}{4}m + \log(m - n/3) \qquad (2.8)$$

where $A$ is a constant depending on $n$. This is the GR law analog for the model which clearly holds for the high magnitude end of the distribution. As we go to the low magnitude end there is a cut-off at $m = n/3$ near which the distribution falls off rapidly. Our model very naturally brings out the low magnitude roll off in GR statistics which has been generally ascribed purely to incomplete reporting (with respect to total number of occurrences) of low magnitude earthquakes [30].

Now the so-called '$b$-value' from our theoretical distribution is $3/4$ and not unity as generally reported. The value $3/4$ arises out of the fact that we have constructed our Cantor set by the middle third removal procedure. In fact for a Cantor set with dimension $\log(q-1)/\log(q)$ the exponent would be $q/(q+1)$. This of course would be effectively unity for higher dimensional Cantor sets. Now this means that there will be region to region variation in the '$b$-value' and this is unique in the sense that most theoretical models [4, 7] give universal values for the exponent. In practice the '$b$-value' has shown some variability from unity. The '$b$-value' generally varies from 0.5 to 1.5 depending on the tectonic setting, tectonic stress and the magnitude ranges [31, 32] but normally comes close to 1 for seismically active regions. The Gutenberg–Richter power law relation holds good for aftershock sequences also which is really what our model describes [33, 34]. In our model however the range is from slightly smaller than 0.75 to 1 (the lower bound on the exponent is smaller than 0.75 in practice as $m^2/n$ is slightly smaller than $m$ in reality).

The constant $A$ in (2.8) is dependent on the generation number $n$ and the value of $n$ determines, for a given similarity dimension, the seismicity in our model i.e. the number of earthquakes increases with increasing $n$. Mathematically, $A$ is equivalent to the constant '$a$' in the GR law. It is notable that in the GR law too '$a$' characterizes seismic activity. So $A$ is a reasonable proxy for the '$a$' value in GR law. Fig. 2.2 shows the GR law plot from the model for generations 8 and 9. The values obtained for the exponent (the '$b$-value') are also indicated in the plot. The values obtained by fitting ($b=0.74$ for $n=8$ and $b=0.71$ for $n=9$) support our analysis presented above. The low magnitude roll-offs are also quite conspicuous for both $n=8$ and $n=9$. Comparison with the frequency-magnitude plot for Sumatra shown in later in Fig. 3.1 clearly brings home the similarities between our theoretical distribution and the form observed in nature.

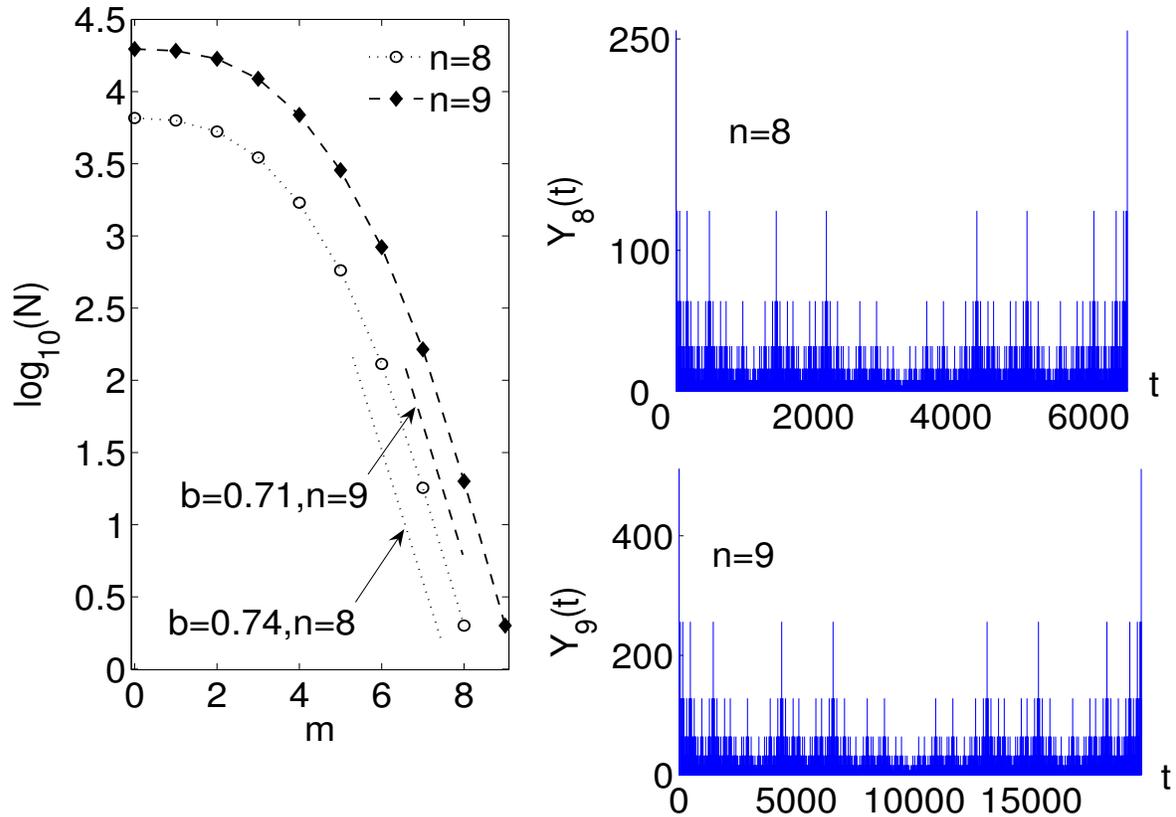

*Fig. 2.2 The Frequency-Magnitude (FM) plots ($log_{10}N(m)$ vs. Magnitude, N being the number of earthquakes with magnitude greater than or equal to magnitude m) for the model for generations 8 and 9. The overlap time series are also shown on the right for the respective generations. The low-magnitude roll-off is evident for both the generations. The lines are drawn as visual aid to understand the linear trend. The 'b' values were obtained by fitiing a linear polynomial to the data. The values of the exponents are the slopes of the indicated straight lines. The 'b' values thus obtained are also indicated for each of the generations.*

## 2.4 The Omori Law

Previously a theoretical study derived the Omori formula from a preliminary statistical model where aftershocks are produced by a random walk on a pre-existing fracture system [35]. The derived result shows a direct connection between $p$ and the fractal dimension of the pre-existing fracture system. This study showed that the fractal properties of aftershocks are determined by the fractal geometry of the pre-existing fracture system. The Omori law comes out very naturally from our fractal overlap statistics as well.

Physically, our model corresponds to an aftershock sequence for a mainshock of magnitude $n$. So it is of inherent interest to check for the Omori Law in our model by studying the temporal distribution of these synthetic aftershocks. The time series of overlap magnitudes in our model has built-in power law behaviour. The entire magnitude-time sequence is a nested structure of geometric progressions as pointed out earlier. This makes it difficult to enumerate an exact value of the exponent $p$ in general. But there is, however, departure from this in two limiting cases. Omori Law in practice gives specific value of $p$ for a given magnitude threshold. We observe that for any generation, when the threshold is the minimum overlap magnitude 1 in our model, the $p$ value is 0. This is because by the virtue of the assumption of uniform velocity there is an aftershock at every time step. A very interesting fact is however unearthed on putting the magnitude threshold at the second highest possible value $n$-1 (that means we are considering aftershocks only of magnitude $n$-1 and higher). Now the times of occurrences of aftershocks of magnitude $n$-1 are at each value of $t = 2.3^{r_1}$ where $r_1$ varies from 0 to $n$-1 (as given in (2.3)). Therefore when the lower magnitude threshold is $n$-1 we have, not considering the constant prefactor 2, consecutive aftershocks occurring at times which follow this geometric progression (2.3) with common ratio 3 (that is if at any $t$ there is an aftershock of magnitude $n$-$1$ then at $3t$ the next aftershock will occur and at $3^2 t$ the one after that and so on). This gives the general rule $N(3t) = N(t) + 1$ leading to:

$$N(t) = \log_3 t \qquad (2.9)$$

where $N(t)$ is the cumulative number of aftershocks (of magnitude $m \geq n$-1 for a mainshock of magnitude $m = n$). Integration of the Omori relation gives, $N(t) = t^{1-p}$.

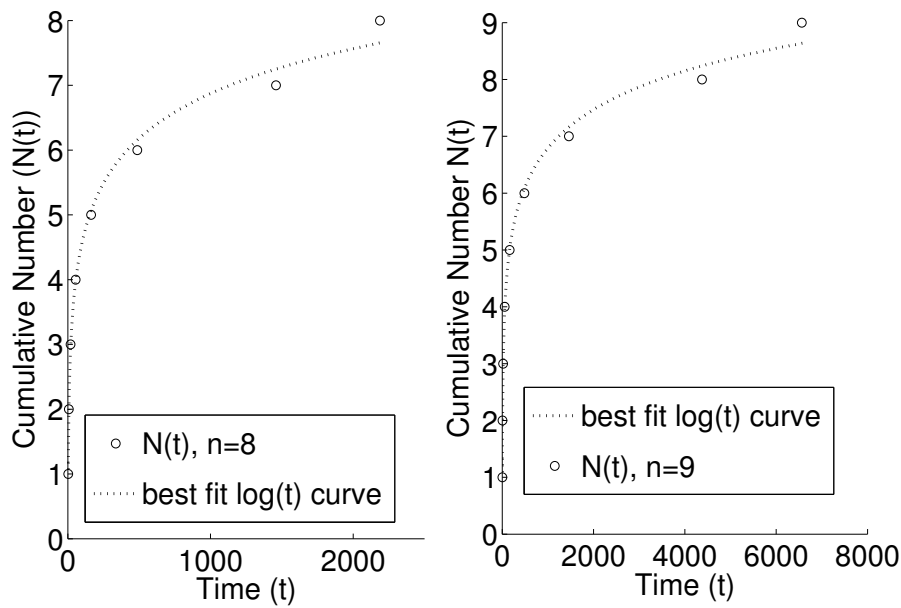

*Fig. 2.3 Omori Law from the model for generations 8 and 9 respectively. Dashed lines show best logarithmic fits. Plots are for N(t) vs. t, N(t) being cumulative number of aftershocks at time t where t is the time since the mainshock. Time parameter for the model being as defined in the text i.e. unit time for a step of size $3^{-n}$.*

From (2.9) this gives us $p \sim 1$ which is the traditional Omori exponent value. The model therefore gives a range of $p$ values from 0 to 1 which systematically increases within the range with increasing threshold. Fig. 2.3 shows the plots (from the model) for cumulative number of aftershocks $N(t)$ of magnitude greater than $2^{n-2}$ versus $t$ for $n=8$ and $n=9$.

The fact that the Omori exponent $p$ is not universally unity is a very well documented fact and some workers have reported variability in $p$ from 0.5 to 2.5 [36]. But for seismically active zones $p$ is generally close to unity. This variability is present in our model too. The variability in $p$ in our model apparently stems from implementing different magnitude thresholds.

But, there is a deeper analogy with the real world. The magnitude threshold for Omori Law calculations is always put above the completeness magnitude. Completeness magnitude is that magnitude below which the frequency-magnitude statistic rolls-off from the GR like power law, that is, the number of earthquakes is not exhaustively recorded below this magnitude and this is the reason for the roll-off. In other words the complete record of earthquakes below this magnitude is not available in the sense that the frequency level below this magnitude is less than what really should be according to the GR law. For real earthquakes Omori Law exponents are calculated only in the power law region of the magnitude scale. In our model such a roll off occurs naturally. The roll-off occurs at approximately below a magnitude $n/3$ as discussed in sub-section 2.3. A meaningful comparison with the Omori statistics for real data sets can be done only for the power law region and that means our threshold can be no smaller than $n/3$. This implies that the $p$ exponent can never be observed to be zero. And for a higher generation or a higher dimension fractal, at the same magnitude cut-off, the $p$ exponent will be higher than a lower dimension or lower generation fractal. Values of $p$ closer to unity will be seen as we take up Cantor sets of progressively higher dimensions and/or generations at the magnitude cut-off $n/3$. The higher the generation and/or dimension of the fractal we consider the higher will be the mainshock magnitude and more number of aftershocks will be observed in the model. Thus the seismic activity will increase. At the same time the exponent $p$ will yield values closer to unity even at magnitude cut-offs lower than $m = (n-1)$. Thus for seismically active zones $p$ values will be closer to unity. This is analogous to what is observed in nature.

## 2.5 Temporal distribution of magnitudes of an aftershock sequence

There is another very important observation that comes out from the model. If we evalu-ate the time cumulant of magnitude, i.e. $\int_0^t m(t')dt'$ where $t$ is the time since the mainshock and $m(t)$ is the magnitude at $t$, it comes out to be a remarkable straight line. In other words:

$$Q(t) = \int_0^t m(t')dt' = St \qquad (2.10)$$

where $S$ is the slope of the straight line. This temporal distribution of the $Q(t)$ statistic is very significant. The slope $S$ is a function of both the generation number as well as the dimensionality of the Cantor set. It is however quite difficult to enumerate the slope exactly due to the presence of the nested geometric progressions in the time series as stated earlier but an approximate estimate of the slope is given by

$$S_n^q = \left(\frac{q-1}{q}\right)\frac{n}{2} \qquad (2.11)$$

for the model where the Cantor set has been formed by removing the middle block from $q$ blocks and the generation number is $n$. Now the important fact coming out of (2.11) is the dependence of $S$ on both the dimension and the generation number of the model. The model predicts that the slope $S$ for real aftershock sequences would be fault dependent as we expect the generation number and/or the dimension of the fractals involved to vary from fault to fault. Thus in a sense, the slope $S$ is a kind of a 'fractal fingerprint' of the fault zone. The slope is a very characteristic local feature of the aftershock sequence and hence of particular interest as a diagnostic feature of aftershock sequences. In effect this provides us a new approach in analyzing the temporal behaviour of aftershock sequences from which we can, at least from the model, clearly extract information about the fault geometry. Fig. 2.4 shows such $Q(t)$ vs. $t$ plots for the model for $n = 7, 8, 9$ respectively. From the figure one can clearly see the increase in slope with successive increases in generation number $n$. The increase in generation number is something that we expect more commonly in an active seismic zone. This can take place due to re-rupturing of an existing rupture zone. Such re-rupturing has been reported very often and happens when an earthquake occurs at or near the hypocenter of a previous large earthquake

(hypocenter is the assumed point from which seismic waves emanate) years afterwards. We discuss such an event and the resultant $Q(t)$ versus $t$ plot in sub-section 3.3.

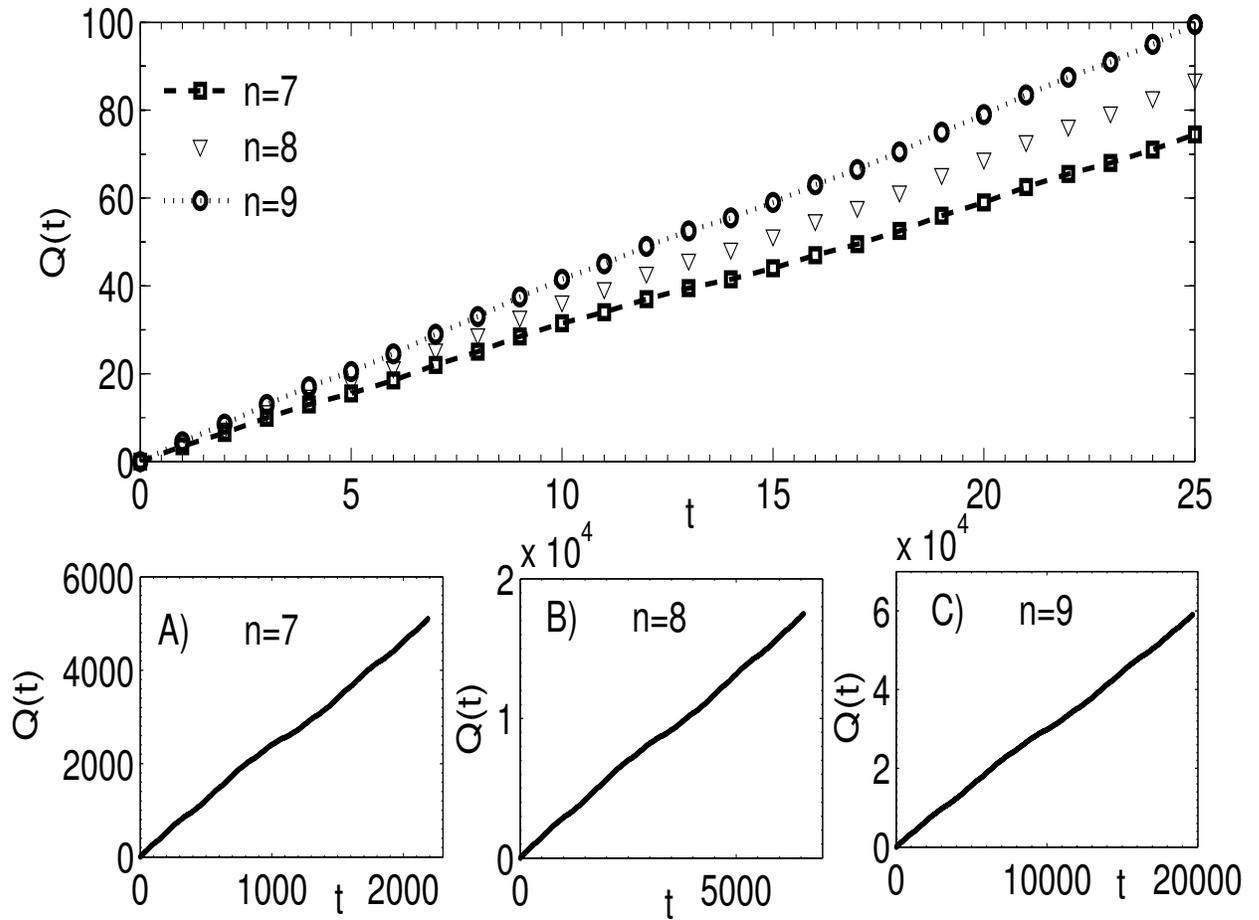

*Fig. 2.4 The Q(t) vs. t statistic for the model for generations 7, 8 and 9. At the top plots for all the three generations (for the first 25 time steps) are shown together to show the increase in slope with increase in generation number n. At the bottom plots A), B) and C) show the entire Q(t) time series for generations 7, 8 and 9 respectively.*

## 3 Comparison with observations:

## 3.1 The Gutenberg Richter law

In Fig. 3.1 we have considered the frequency-magnitude distribution for two real aftershock sequences to compare our theoretical formulation (see sub-section 2.3) with real earthquake data. The data sets considered were 1) The 2004 Sumatra earthquake aftershock sequence (26/12/2004, $M_w$ = 9.0, Epicenter latitude: 3.30$^o$, Epicenter longitude: 95.98$^o$, source catalog: NEIC (PDE) catalog [37]) and 2) The 1995 Kobe earthquake aftershock sequence (17/01/1995, $M_{JMA}$ = 7.2, Epicenter latitude: 34.6$^o$, Epicenter longitude: 135.0$^o$, source catalog: JUNEC catalog [38]). Aftershocks of a major event were considered to be events within a given region, geographically defined as boxes or polygons constrained by suitable latitudes and longitudes, and the magnitudes were recorded over a length of time (of the order of a year or more) over which the region has not yet relaxed to its background seismicity (tentatively within the first 1000 days). Now one point needs to be made clear with respect to the Sumatra dataset. The dataset was inhomogeneous in the sense that it reported earthquake magnitudes in different magnitude units. So we had to convert all the magnitudes reported to one uniform magnitude scale using inter-magnitude conversion relationships. We chose the uniform magnitude scale for our work to be the moment magnitude $M_W$ as defined in [39]. For the Sumatra event we used the conversion relationships used in [40]. These relationships were specifically designed for the aftershock sequence of the Sumatra event extracted from the PDE catalog and hence serve our purpose. The fact that the conversion relationships were designed for nearly the same dataset as we have used here is important as such conversion models are in general regressional models and hence their use in our work is validated by the fact that here we use them on the same population for which they were originally designed. But errors in magnitude reporting as well as those induced due to magnitude conversions can severely affect the estimation of the GR law exponent. These errors have been discussed in a bit more detail in sub-section 3.3. As we remarked earlier, the frequency-magnitude plot for the Sumatra aftershock sequence clearly shows the roll off from GR statistics at the low magnitude end. This is similar visually to the roll off observed in our model (see sub-section 2.3 and Fig. 2.2). But the Kobe sequence does not show any such clear roll-off.

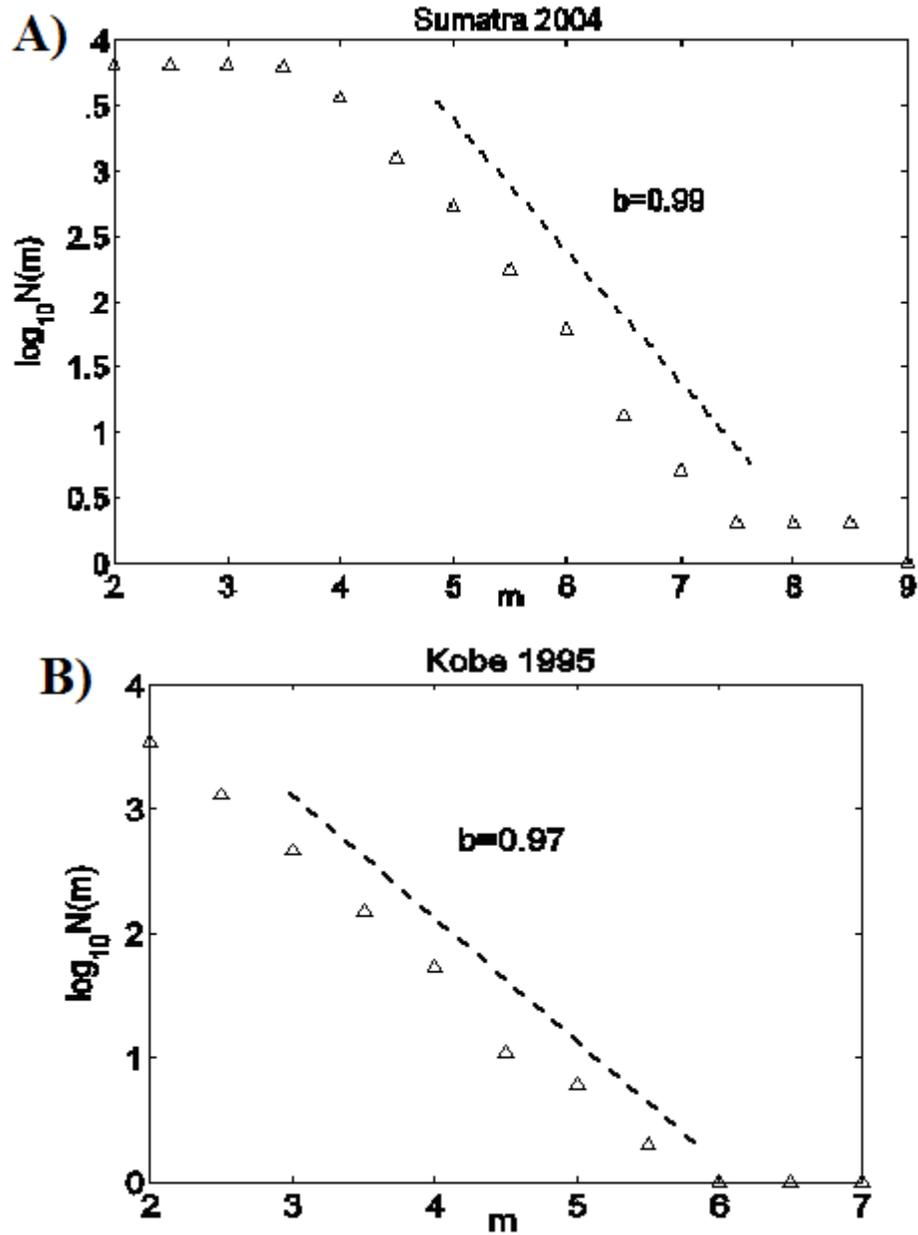

*Fig. 3.1 GR or Frequency-Magnitude distributions for the aftershock sequences described in the text A) the 2004 Sumatra earthquake and B) the 1995 Kobe earthquake. We clearly see the lower magnitude roll-off from the power law distribution in Sumatra. N(m) represents number of earthquakes with magnitude greater than or equal to m, m represents magnitude.*

## 3.2 The Omori law

Our model shows that the Omori exponent $p$ (see equation (1.3)) increases with increase in the lower magnitude threshold. We tried to check for this trend of increase in $p$ with increase in lower magnitude threshold for three real aftershock data sets. The aftershock sequences chosen were 1) 1989 Loma Prieta earthquake aftershock sequence (18/10/1989, $M_w$ = 7.1, Epicenter latitude: 37.0°, Epicenter longitude: -121.88°, source catalog: [41]); 2) 1999 Chamoli earthquake aftershock sequence (29/03/1999, $M_S$ = 6.6, Epicenter latitude: 30.51°, Epicenter longitude: 79.40°, source catalog: WIHG catalog [42]) and 3) 2004 Sumatra earthquake aftershock sequence described before. The results are given as $\log n(t)$ vs. $\log t$ plots in Fig. 3.2 where the cut-off thresholds are denoted as $M_c$ and the $p$ values are indicated. Here $n(t)$ denotes number of aftershocks per unit time and $t$ denotes time since the mainshock in days. As is evident from Fig. 3.2, the increase in $p$ with increase in $M_c$ is clearly seen in Chamoli and in Sumatra. However in Loma Prieta, which is a very well characterized data set, the same trend is not seen. The reason for widely different values of $M_c$ for the three data sets is that the completeness level (as explained earlier meaningful analysis can only be done above the completeness magnitude) for the three catalogs are very much different mainly due to the nature of the seismic networks implemented.

## 3.3 The temporal distribution of aftershock magnitudes:

The linearity of the $Q(t)$ statistic was checked with magnitude-time sequences for real aftershock sequences. We first collected the aftershock magnitude-time sequences $m(t)$ of eleven major earthquakes from different catalogs from different geographical regions of the world. The earthquakes were selected carefully from all over the globe to ensure that no regional bias was introduced due to the choice of a specific catalog or a specific geological setting. We also intentionally selected some multiple events in the same geological region on a) different fault zones b) the same fault zone at a different time. We then evaluated the cumulative integral $Q(t)$ of the aftershock magnitudes over time. A trapezoidal rule was used to evaluate $Q(t)$; here $t$ denotes the time since the main shock. The various events for which we carried out our analyses are described in Table 3.1.

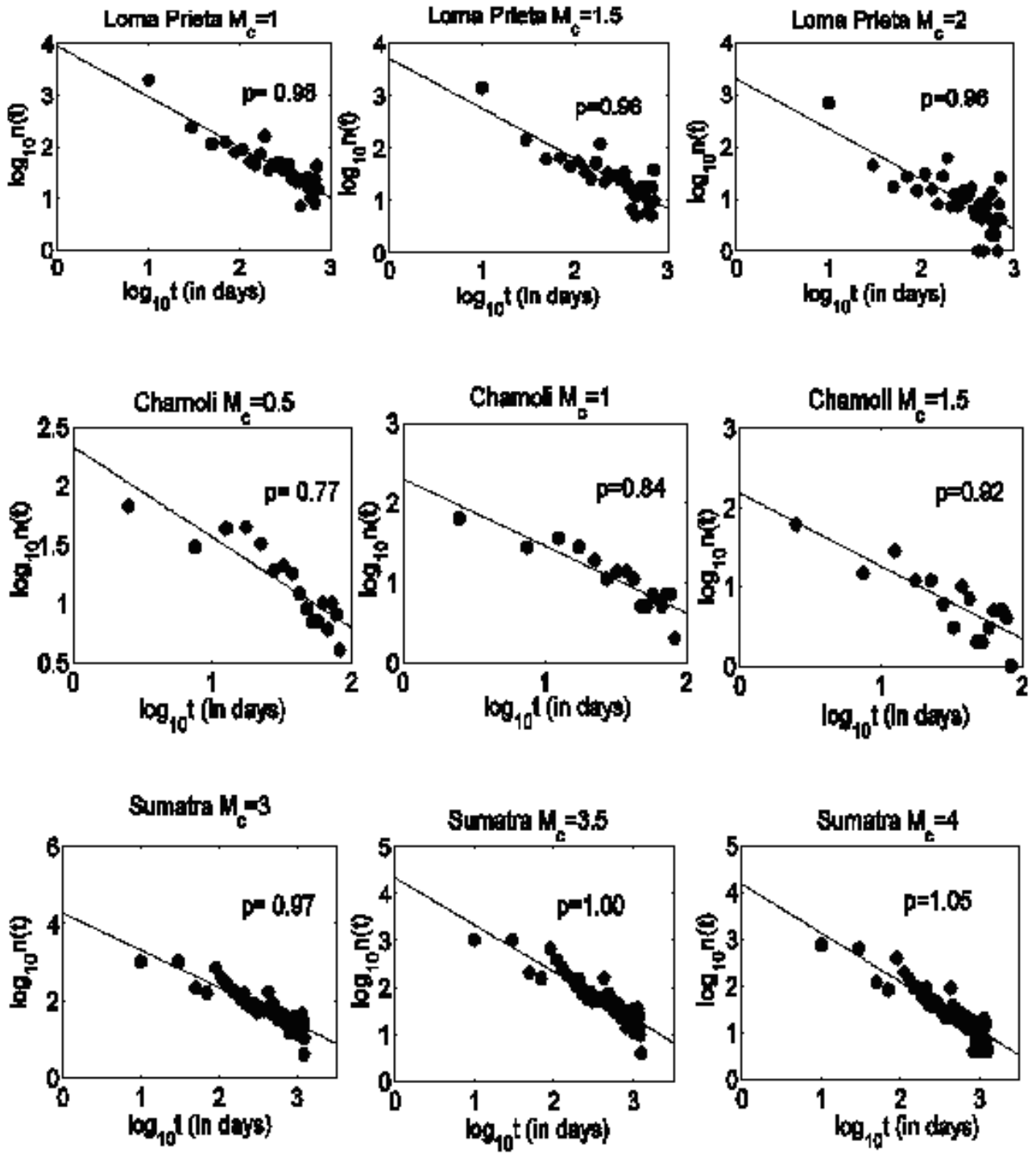

*Fig. 3.2 The plots for number of aftershocks per unit time n(t) vs. time since the mainshock in days t for the Loma Prieta, Chamoli and Sumatra data sets. The cut-off magnitudes $M_c$ are indicated in each plot title. The corresponding p values are shown within the plots. The solid lines give the linear fits to the data with slope p.*

| Event Name | $S_1$ | $S_2$ | $S_{loc.1}$ ($\sigma_1$) | $S_{loc.2}$ ($\sigma_2$) | Event Description (dd/mm/yyyy, magnitude, epicenter lat., epicenter long.) |
|---|---|---|---|---|---|
| Loma Prieta | 1.74 | - | 1.85 (0.27) | - | 1989 Loma Prieta earthquake (18/10/1989, $M_w$ = 7.1, 37.0°, -121.88°). Source: [41] |
| Kobe | 2.39 | - | 2.43 (0.13) | - | 1995 Kobe earthquake (17/01/1995, $M_{JMA}$ = 7.2, 34.6°,135.0°). Source: JUNEC catalog [38] |
| Sumatra | 4.47 | 4.07 | 4.55 (0.18) | 4.16 (0.21) | 2004 Sumatra earthquake (26/12/2004, $M_w$ = 9.0, 3.30°, 95.98°). Source: NEIC (PDE) catalog [37] |
| Muzaffarbad | 3.93 | 4.00 | 4.03 (0.23) | 4.10 (0.24) | Muzaffarabad (Kashmir, North India) earthquake of 2005 (08/10/2005, $M_S$ = 7.7, 34.52°, 73.58°). Source: NEIC (PDE) catalog [37] |
| Chamoli | 1.95 | - | 2.07 (0.36) | - | Chamoli earthquake (29/03/1999, $M_S$ = 6.6, 30.51°, 79.40°). Source: WIHG catalog [41] |
| Bam | 3.33 | - | 3.32 (0.17) | - | The Bam earthquake (26/12/2003, $M_S$ = 6.8, 29.00°, 58.31°). Source: IIEES catalog [65] |
| Zarand | 3.40 | - | 3.36 (0.12) | - | The Zarand earthquake (22/02/2005, $M_S$ = 6.5, 30.80°, 56.76°). Source: IIEES catalog [65] |
| Alaska 1 | 3.02 | - | 3.20 (0.26) | - | Denali fault earthquake in Alaska (03/11/2003, $M_S$=8.5, 63.52°, -147.44°). Source: NEIC(PDE) catalog [37] |
| Alaska 2 | 3.39/4.05 | - | 3.53/4.06 (0.34/0.17) | - | Rat Islands, Aleutian Islands earthquake in Alaska (17/11/2003, $M_W$ = 7.8, 51.15°, 178.65°). Source: NEIC (PDE) catalog [37] |
| Taiwan 1 | 4.11 | - | 4.10 (0.16) | - | Taiwan earthquake (31/03/2002, $M_W$ = 7.1, 24.13°, 121.19°). Source: BATS CMT catalog [66] |
| Taiwan 2 | 4.26 | - | 4.25 (0.29) | - | Taiwan earthquake (26/12/2006, $M_W$ = 6.7, 21.89°, 120.56°). Source: BATS CMT catalog [66] |
| Honshu 1 | 4.34 | - | 4.42 (0.21) | - | Honshu earthquake (31/10/2003, $M_W$ = 7.0, 37.81°, 142.62°). Source: NEIC (PDE) catalog [37] |
| Honshu 2 | 4.32 | - | 4.37 (0.19) | - | Honshu earthquake (16/08/2005, $M_W$ = 7.2, 38.28°, 142.04°). Source: NEIC (PDE) catalog [37] |

*Table 3.1 Event names are used to refer to respective sequences in the text. The event tags correspond to those in the plot in Fig. 3.3. $S_1$ corresponds to the slope of the linear fit with the raw data while $S_2$ corresponds to the linear fit with the homogenized data set. The additional subscript loc. for columns 6 and 7 give the averages of the local slopes for the raw and homogenized data respectively. For the Alaska 2 aftershock sequence, the slope changed midway (see Fig. 3.4) and the two slopes depict the slope for the earlier part and after the slash the slope for the later part for both $S_1$ and $S_{loc.1}$. $\sigma_1$ and $\sigma_2$ respectively denote the standard deviations of the local in time slope versus time distributions for the unconverted and converted magnitudes. These are reported in parentheses along with $S_{loc.1}$ and $S_{loc.2}$ respectively.*

The important limitation of our analysis, while evaluating the aforementioned integrals, is the fact that more often than not most catalogs which give the most exhaustive list of aftershocks report the various events in different magnitude scales. This again warrants the need for using conversion relationships to convert the various magnitude scales to a uniform scale (as done previously for the GR law). This, wherever we have inhomogeneous catalogs, we have chosen to be $M_w$ (once again like we did in the case of the GR law), the moment magnitude as defined by Kanamori [39]. To this end we have used well defined and previously employed conversion relationships specifically designed for the sequences herein. The datasets extracted from the NEIC (PDE) catalog are all inhomogeneous with respect to the magnitude scales used to report the various events. The PDE listing was used to obtain the aftershock sequences of the Sumatra, Muzaffarabad, Alaska and Honshu events (see Table 3.1). For the Sumatra aftershock sequence we again used the conversion relationships used in [40]. For the Muzaffarabad sequence we used conversion relations given in [43] which were designed specifically for the region and is based on the same NEIC (PDE) listing. Table 3.1 describes the names that we have used in the text subsequently to refer to the respective aftershock sequences. For the aftershock sequences Alaska 1, Alaska 2, Honshu 1 and Honshu 2 we could not obtain valid conversion relationships and hence for these datasets we have used the inhomogeneous catalog in its raw form supplemented by a less extensive homogeneous aftershock magnitude listing for the same magnitude-time sequence. This means if the most number of events were reported in say the $m_b$ scale (the body wave magnitude scale), then we extracted the list of only these events and evaluated $Q(t)$ for these events separately.

      Our analyses indicate a clear linear relationship $Q(t) = St$ where $S$ denotes the slope. The fitted slopes from the unconverted magnitude datasets are represented by $S_1$ in Table 3.1. The converted magnitudes (with the conversion relationships mentioned above) were then used to re-evaluate $Q(t)$ and the fitted slopes are represented by $S_2$. We then additionally computed the local (in time) slopes for $Q(t)$ for overlapping time segments of the $Q(t)$ statistic. The lengths of these segments were selected in accordance with the size of the respective datasets. The mean of this temporal distribution of slope is represented by $S_{loc.1}$ in Table 3.1 for the unconverted sequence and the standard deviation is denoted

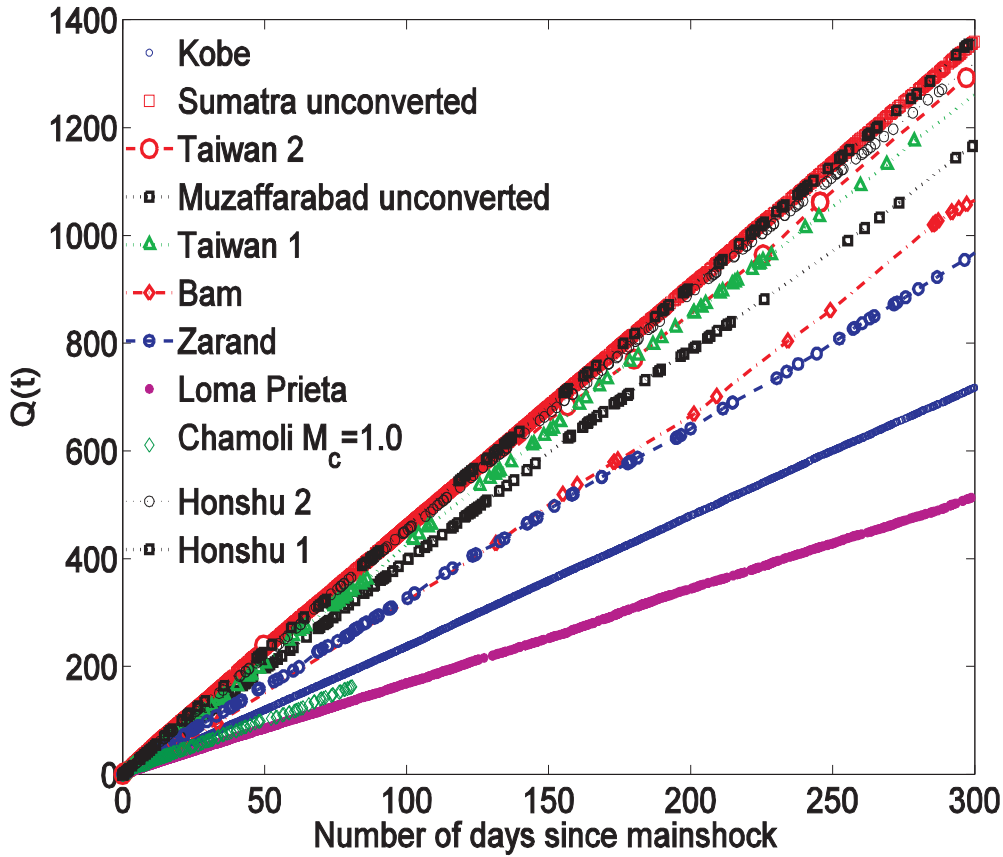

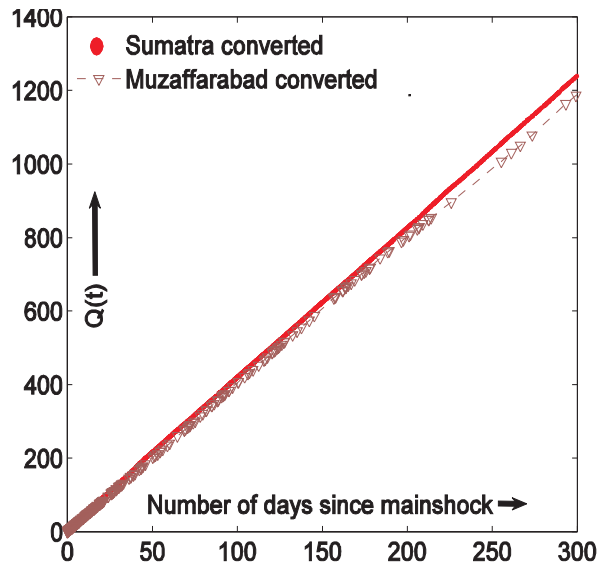

*Fig. 3.3 Top: Plots of time cumulant of magnitude Q(t) vs. t (in number of days since the mainshock) for the datasets described in the text and in Table 2.1 for the first 300 days. Bottom: Plots of Q(t) vs. t for the Sumatra and Muzaffarabad sequences after conversion of magnitudes according to [40] and [43] respectively.*

by $\sigma_1$. Similarly, for the converted datasets the mean local slope is denoted by $S_{loc.2}$ in Table 3.1 and the standard deviation is represented by $\sigma_2$. The plots are shown in Fig. 3.3 (with the exception of the two Alaska events).

The results of our analysis mentioned in Table 3.1 and the plots in Fig. 3.3 point clearly to the linear variations mentioned above. The straight line (fit) retains this slope for years. Also, the slope changes once we compare the statistic $Q(t)$ for different fault zones. This indicates that $S$ is characteristic of the fault zone. This was further checked by integrating from anywhere in the time series (i.e. shifting our $t = 0$ to any randomly chosen aftershock) after the mainshock. This was done for both the raw as well as the homogenized data. The slopes for such plots were found to be within 2% variability, with respect to the integral evaluated since the mainshock, for all the data sets analyzed. Also, the 2% variability in slope is clearly within the error bounds induced by the data sets. A wide variety of events can lead to systematic errors in the reported magnitudes (events as varied as a change in instrumental calibration to addition or removal of seismograph stations) and such systematic errors can be very large going up to as much as 0.5 magnitude units [44]. Such errors would set the eventual error bound for the slope as the errors due to fitting are much smaller as mentioned already. Additionally, the conversion relationships themselves induce some errors in the magnitudes. This can also lead to systematic errors in the slope estimate. With the available catalogs, the errors in slope estimation would be thus about 6-10% [44, 45]. But the effect of changing the lower magnitude threshold has some effect on the magnitude of the slope. Systematically increasing the lower cut-off of the magnitudes considered systematically increases the slope. But this has a very simple explanation. We have observed that the magnitude-time sequences for any real earthquake aftershock sequence are such that the slope of the statistic $Q(t)$ gives, approximately, the average magnitude of the sequence. This implies that a large mainshock with large aftershock magnitudes will have a large slope while a similar mainshock magnitude with a large number of smaller magnitude aftershocks will have a smaller slope. So changing the lower cut off for magnitude would change the slope as it would affect the averaging procedure that we are mathematically carrying out. It is very clear that the average magnitude of a given aftershock sequence would depend on the stresses involved and the asperity distribution on the specific fault zone. This is the basis of our claim that the slope is characteristic of the fault zone. Thus for a given catalog with given completeness level the slope is characteristic of the fault zone. The change in slope due to changing the lower cut-off of magnitude was however observed to be within our 10% error bound when the escalated completeness magnitude was far less (about a magnitude order less) than the slope obtained due to fitting. To illustrate the characteristic feature further we draw attention to the two Taiwan sequences, Taiwan 1 and 2. Both of these took place on the Eurasian and Philippines plate boundary. While Taiwan 1 took place in a region where the convergence of the plates is compensated by crustal shortening, Taiwan 2 took place in a region where the oceanic Eurasian plate is subducting. But the rupture zones are both on the same plate boundary and later events in the Taiwan 2 sequence, including a very large event approximately 8 minutes after the mainshock, may have occurred within the compressional regime. The style of faulting for the subsequent large event was consistent with the tectonism observed in the rupture zone of Taiwan 1 [46]. It is only natural that the two corresponding slopes would be nearly identical in view of our proposed error bounds due to the geological similarities and precisely similar tectonism. Again the two Honshu events stand in strong support of our proposition that the slope is characteristic of the fault zone. The locations and focal-mechanisms of both these earthquakes imply that they occurred as the result of thrust-faulting on the plate interface between the overriding Okhotsk plate (between the Pacific Ocean and the Eurasian landmass) and the subducting Pacific plate [47]. The Pacific plate is moving west-northwest at a rate of about 83 mm per year relative to the Okhotsk plate in this region and this regions has very high seismicity. Again occurrence of two separate events in the same fault zone and tectonic regime give us the same slope (within proposed error bounds). In Iran though, the Bam and Zarand earthquakes took place on two different faults belonging to a highly developed fault system. The Bam event occurred on the Bam fault whereas the Zarand event took place in close proximity of a previous event on the Gowk system (1981 July 28, Sirch earthquake $M_W = 7.1$) at a distance of about 60 km from the northern extremity of the rupture zone of the Sirch event [48]. But still the slopes were found to be the same. The slope does not change with unusually large aftershocks in the sequence e.g. the Sumatra sequence had a few very large aftershocks including one great earthquake on 28[th] March 2005 ($M_W = 8.7$) which occurred about 150 km SE of the earlier giant earthquake epicenter ($M_W$ = 9.3) of 26$^{th}$ December 2004. This further reveals the characteristic nature of the slope.

In view of the above discussion, we further strengthen our claims using the results for the two sequences obtained in Alaska. Alaska 1 was an event on the inland Denali fault and the $Q(t)$ statistic gives a slope $S_1$ = 3.02. The localized slope estimate was $S_{loc.1}$ = 3.24. We did not find a good conversion relationship for this sequence and instead used the most numerous magnitude type in the sequence which was the local magnitude $M_L$. This gave us a homogeneous listing of events and we recalculated the slope to obtain 3.08 for $S_1$ and 3.23 for $S_{loc.1}$. One more aspect came out during the analysis of the Alaska 1 dataset. The first shock considered here was not the Denali fault mainshock but a previous shock in the same region. This was done because this event is a very well established foreshock of the Denali fault event. This shows something very important. For events on the same fault system the slope is the same and hence it holds for foreshocks too. This linearity and constancy in slope are very local and the slope is the true identity of the rupture zone. This claim is further strengthened on analysis of the Alaska 2 aftershock sequence (see Fig. 3.4). Here, the slope of the $Q(t)$ vs. $t$ curve increases after about 800 days of the main event. We first need to understand the tectonics of the region where the events in Alaska we have considered have occurred [49]. One of the most significant events of the last century, the 1965 Mw 8.7 Rat Islands earthquake, ruptured a ~600 km-long portion of the plate boundary to the west of the Amchitka Island. In the November 17$^{th}$, 2003 M7.7 earthquake, the main shock or the first shock in the sequence we chose, the easternmost part of the 1965 zone failed again. On June 14, 2005, a series of moderate to strong earthquakes occurred in the Rat Islands region of the Aleutian Islands. The sequence started with a M5.2 event at 08:03 UTC and the largest event of M6.8 followed 9 hours later (at 17:10 UTC). The largest earthquake was situated 49 kilometers (31 miles) south-southeast of Amchitka. This new sequence of earthquakes re-ruptured the easternmost end of the 1965 rupture zone. This re-rupturing is the reason, we believe, for the increase in slope. The re-rupturing process meant that the earlier asperity distributions were changed and hence the region underwent a marked change in its seismicity pattern. In general re-rupturing of a fault would imply an increase in generation number. A change in dimension is quite unlikely. By (2.11) it is easy to show that this would increase the slope *S* of the *Q(t)* statistic. In fact this kind of scenario is very helpful in estimating fractal properties of the fault zone. Under the assumption that the dimension has not changed we can estimate the change in generation number. But whether that estimate is accurate enough depends on the accuracy of the data set. Here, as mentioned earlier, we have an inhomogeneous magnitude reporting and therefore the estimates might not be reliable enough. In view of the fact that the slope changed with the completeness magnitude of the catalog, it would have been reasonable to put forth the conjecture that the slope was really only a function of the various completeness magnitude cut-offs that the various catalogs have for the various geographical regions. But here, the catalog being the same with the same completeness level throughout, the increase in slope clearly establishes that the slope is characteristic of the specific aftershock sequence and is not merely a regional feature or an artifact of the magnitude listings in the given catalog. The various slope measures for the Alaska 2 sequence are given in Fig. 3.4 for the total inhomogeneous event listing i.e. $S_1 = 3.39$ for the earlier half and $S_1 = 4.05$ for the later half. We further observed $S_{loc.1} = 3.39$ for the earlier half and $S_{loc.1} = 4.05$ for the later half. We extracted the homogeneous $M_L$ listing for this sequence too and obtained $S_1 = 3.09$ for the earlier half and $S_1 = 3.64$ for the later half and also $S_{loc.1} = 3.09$ for the earlier half and $S_{loc.1} = 3.76$ for the later half. As $M_L$ listing conventionally avoids the larger events, the slopes are reduced for reasons similar to the ones discussed previously.

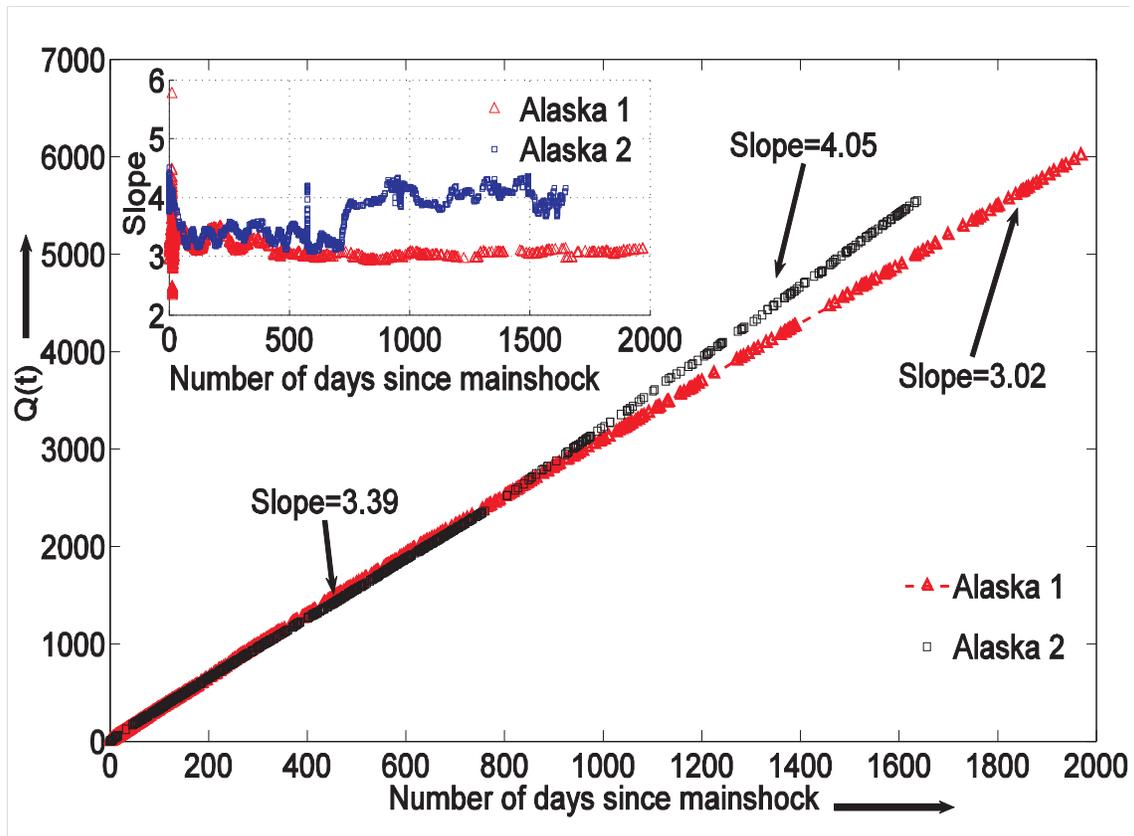

*Fig. 3.4 Plots of Q(t) vs. t for the datasets Alaska1 and Alaska2 described in the text. Note the increase in slope for the Alaska 2 sequence after about 800 days of the 2003 event i.e. approximately at the end of 2005. This is just after the re-rupture of the fault zone. Inset is a plot of the localized slope vs. time curves.*

However such cumulative statistics have already been attempted for the scalar seismic moment or Benioff stresses for aftershock sequences. We did a similar cumulative integral of scalar seismic moment for our sequences in Sumatra and Taiwan (the former was reliably converted to scalar seismic moment in [33] and the BATS CMT catalog for Taiwan was homogeneous and listed only broadband $M_W$ values). The results are shown in the Fig. 3.5. The resultant plots resemble a step function. In [33] the authors have tried to fit a power law and/or linear models piecewise to such data (in their case the cumulative Benioff stress). There seems to be no robust feature to this statistic, i.e. the cumulative moment versus time curve. Such cumulative curves have also been reported for theoretical models such as for the Critical Continuum-State Branching Model of Earthquake Rupture [29]. Precursory accelerating moment release before large earthquakes has been a widely discussed phenomenon since recent years, being regarded as observational evidence for the controversial critical-point-like model of earthquake generation [50, 51]. Another useful property of such seismic moment cumulants is that they help in monitoring the stress release modes for a given region and hence allow for discussions on the type of mechanisms underlying earthquake occurrences [52].

Our scheme on the other hand addresses a different issue altogether. It gives a very robust and well characterized feature of the sequence instead and the trend of $Q(t)$ versus $t$ is precisely linear. The linearity of the $Q(t)$ statistic has an immediate consequence. Given any aftershock sequence we have this simple recipe. Compute $Q(t)$ for the first few shocks. Then we can assume the magnitude of the very next aftershock to obtain its time of occurrence by linear extrapolation. This means we know that at what time to expect an aftershock of a given magnitude. If we have some presumption of the next shock magnitude then this is especially helpful. We did this for all the sequences and found very good results early in the series. Of course here we knew the magnitude of the next shock. But estimates became erroneous as we extrapolated for later points. This has a very simple solution too. The bad estimates are due to accumulation of error on cumulating large values. This can be easily circumvented as, as discussed earlier, the $Q(t)$ statistic is linear irrespective of which shock in the sequence you start integrating from. So we re-evaluated the statistic every ten to fifteen points and then obtained time estimates as close to the true time of occurrence as a few minutes. This faculty is afforded only by a linear model as extrapolation truly makes sense only when the model is linear. In this aspect the $Q(t)$ statistic is unrivalled. This gives us a very new way of estimating the time of occurrence of aftershocks of a given magnitude sequence provided that we know the slope of the $Q(t)$ versus $t$ distribution. To really achieve this however we need again very accurate magnitude determination and reporting. With the advent of real time seismic monitoring this scheme might go a long way in providing successful forecasts provided we have some idea of the next imminent aftershock magnitude or at least the order of the next aftershock magnitude.

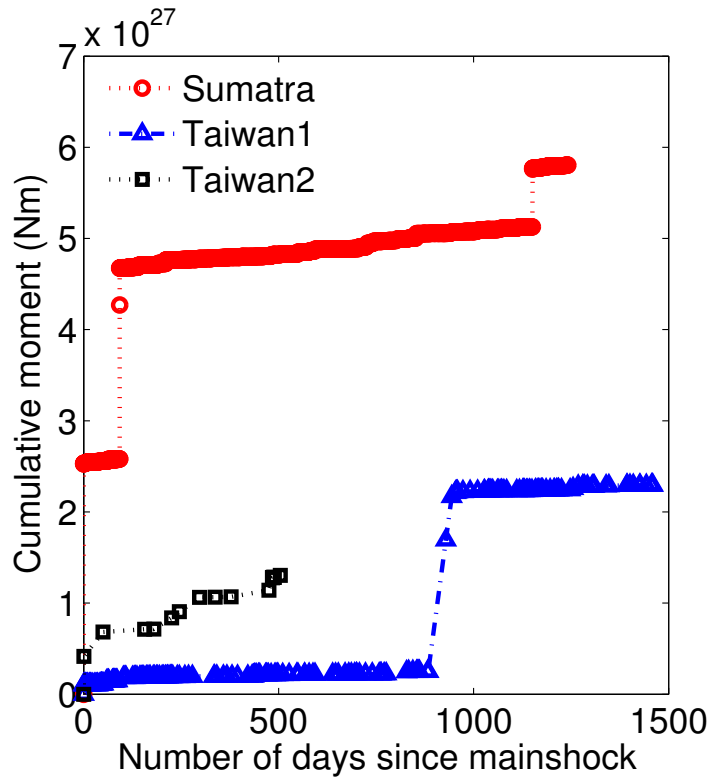

*Fig. 3.5 Plots of cumulative moment versus time since the mainshock for the datasets Sumatra, Taiwan 1 and Taiwan 2. The values for Taiwan 2 depicted here in the plot are 10 times the real values to ensure proper legibility of the figure.*

# 4 Fiber bundle model of earthquakes

In this section we undertake the analysis of earthquake dynamics from the point of view of material properties of deformable materials that break under a critical stress. This approach of modeling is entirely different from the geometric approach discussed till now in Sections 2 and 3. The role of fault surface geometry is not considered as much in this approach and emphasis is laid on the stress states involved in the production of seismic activity. Fibre Bundle Models are a typical example of this class.

Earthquakes can be viewed analogous to the brittle failure of the homogeneous materials. Initially local failure starts with random appearence of damage in the form of uncorrelated microcracks, then coalescence of microcracks at the initiation of global failure and finally there is catastrophic occurence of material spanning crack indicating the global failure. Ideally a single preexisting crack is sufficient to trap the applied stress at the sharpest part, leading to global failure. But the phenomenon of propagation of crack is still not well understood due to the complexities of the stress singularities at the crack tip.

Though failure of a material is a complicated phenomenon, it can be mimicked in fiber bundle model. Many authors have modeled the failure of composite materials using the concept of fiber bundle model [53]. This model is basically a microscopic model cosisting of a large number of fibers (constituting a bundle) having different threshold and subjected to longitudinal stress due to external load .

The fibers are considered to have identical elastic constants until failure, while the individual fibers differ in their failure thresholds, given by a distribution. As the external load is shared equally by the intact fibers at any time $t$, the fibers having their failure threshold below this average load per fiber fail. The load shared by this failed fiber is then equally shared by all the other intact fibers in the GLS (Global Load Sharing) model. This extra redistributed load may induce further failure and thus the avalanches continue. The bundle survives when the average load per intact fibers is below the strength threshold of each of the intact fibers. For strongly bonded fibrous materials, excess stress is redistributed to the neighbouring fibers in local load sharing (LLS) hypothesis.

Fiber bundle model under GLS scheme has power law behavior for the size distributions of avalanches

$$\rho(s_a) \propto s_a^{-\chi}, \qquad (4.1)$$

where $s_a$ is the avalanche size given by the number of broken fibers subsequent to an initial failure due to an increase in the external load. The exponent value $\chi$ depends on the manner in which load is increased. For continuous load increase, $\chi$ asymptotically attains the value $5/2$ [54, 55]; whereas for stepwise increase of load, it attains the value 3 [56]. It is obvious that the stepwise increase of load is more realistic from practical poit of view. One simple derivation of the exponent in latter case for a special threshold distribution is given below. This avalanche distribution (4.1) can be interpreted in the context of earthquakes as Gutenberg-Richter frequency-magnitude law.

If one denotes the fraction of intact fibers at any instant $t$ by $U_t$, then for uniform distribution of fiber strength threshold (with a cut-off renormalized to unity), one can write a simple recurrence relation [56]

$$U_{t+1} = 1 - \frac{\sigma_0}{U_t}. \qquad (4.2)$$

Depending upon the initial stress $\sigma_0$, the dynamics finally terminates, resulting in either partial failure or complete failure of the bundle. Using the condition $U_{t+1} = U_t = U^*(\sigma_0)$ for the fraction of surviving fibers at the end of dynamics, one can have

$$U^*(\sigma_0) = \frac{1}{2} + (\sigma_f - \sigma_0)^{\frac{1}{2}}; \quad \sigma_f = \frac{1}{4}. \qquad (4.3)$$

Here $\sigma_f$ is the critical value of initial stress below and at which the dynamics ends up with non-zero fixed value. At $\sigma_0 = \sigma_f = \frac{1}{4}$ the above recursion relation has the time solution

$$U_t = \frac{1}{2}(1 + \frac{1}{t+1}). \tag{4.4}$$

Pradhan et al. [56] provided a way to determine the distribution function $\rho(s_a)$ for avalanche sizes $s_a$. Load in each fiber is increased steadily by an amount $d\sigma_0$ at each step. Avalanch size $s_a$ can be defined as the number of eventual failures due to the change of $\sigma_0$ by this amount and is given by

$$s_a = \frac{dN_b^*}{d\sigma_0} \sim (\sigma_f - \sigma_0)^{-\frac{1}{2}}; \quad N_b^* = N_0(1 - U^*). \tag{4.5}$$

Now, $\rho(s)\Delta s$ measures $\Delta\sigma_0$, or the number of times $\sigma_0$ to be increased by $d\sigma_0$ amount. Hence, using (4.3) and (4.5) the probability distribution function for avalanche size appears to be

$$\rho(s_a) = \frac{d\sigma_0}{ds_a} \sim s_a^{-\chi}; \quad \chi = 3. \tag{4.6}$$

This is analogous to Gutenberg-Richter law with the exponent value equal to $3$.

At $\sigma_0 = \sigma_f$ the number of surviving fibers is $N^*(\sigma_f) = N_0/2$ and and the system will take infinite time to reach the fixed point at $\sigma_f$. Following (4.4), the time variation of the number of broken fibers $N_b = N_0(1 - U_t)$ at $\sigma_0 = \sigma_f$ becomes

$$\frac{dN_b}{dt} = \frac{N_0}{(1+t)^2}. \tag{4.7}$$

This relation is similar to the form of the modified Omori's law [57]

$$\frac{dN_r}{dt} = \frac{k_1}{(k_2 + t)^p}. \tag{4.8}$$

Here, $N_r$ is the number of aftershocks with magnitude greater than some specific value, $p$ is a positive exponent having value near unity (as in the original Omori law (1.3)), $t$ is the time elapsed after the main shock and $k_1$ and $k_2$ are two constants. The above fiber bundle model calculation of course gives the exponent $p$ to be $2$.

Turcotte and his co-workers [58, 59] utilized an another variant of fiber-bundle model along with continuum-damage model to reproduce the same modified Omori's law.

# 5 Summary and discussions

We have presented here some new results that have come out from our analysis of the Two Fractal Overlap model. The model is based on the fact that fault surfaces, both fresh and weathered, exhibit a fractal topography (sub-section 1.2). The model captures the 'stick-slip' dynamics of overlapping fractal surfaces by using regular middle third removal Cantor sets (sub-section 2.1) wherein a Cantor set of a given generation slides over its replica with uniform velocity. The statistical features of the synthetic earthquake time series thus produced are completely analytically tractable (sub-section 2.2). The model, as is evident from our analysis captures both the GR law (sub-section 2.3) and the Omori law (sub-section 2.4). It gives a hitherto unknown statistical feature of the temporal distribution of aftershock magnitudes which we have shown in sub-section 2.5. Moreover, in Section 3 we have shown the proximal correspondence of the values of the model parameters (that is the parameters which describe the statistics of the synthetic seismic activity with the observed values of the statistical parameters (that is the '*b-value*', the constant $a$ in the GR law and the exponent $p$ in Omori's law) describing natural seismicity. The new statistical law discussed in sub-section 2.5 is also very closely followed in nature and this promises to give us important information about the fractal geometry of the faults involved in producing an earthquake and its aftershock sequence. In that sense this new law provides a 'fractal-fingerprinting' of faults.

Then in Section 4 we describe another class of models which describe earthquake dynamics in terms of material properties of deformable materials which break under an applied critical stress. The Fiber Bundle Model, a typical example of this class of models, is discussed. It is shown that GR like and Omori like laws are extractible from such considerations though the values of the parameters of the statistics describing the synthetic seismicity produced in the model are not very close to observed values of the corresponding parameters for naturally occurring earthquake statistics.

Our focus here was on the Two Fractal Overlap Model which is a very simplistic model of earthquakes and do not claim in any way that this is the true scenario that takes place at the geological faults. But the fact that such a simplistic model mimics so much of nature is truly astonishing. The analysis, we reiterate, is one which requires very basic mathematics. The features are strikingly analogous to the real earthquake statistics, they are robust and the variations in the GR and Omori parameters (*b* and *p* respectively) too are very close to what is observed in nature (Section 3). This model clearly shows that much of the statistics of earthquakes can be reproduced under purely geometric considerations of the fault surfaces. Obvious extensions to this work would be to consider overlapping of random Cantor sets or to incorporate different spatial clustering of the Cantor set segments following theories of rock mechanics and fault dynamics. But again, complete analytical tractability would be desirable as only then is the complete understanding of the variations (of the parameters of earthquake statistics) possible. In Appendix C we show that at least the GR law can be numerically extracted from the overlap of random Cantor sets and Sierpinski carpets also. The fractal overlap model opens up a new horizon in earthquake modeling and promises a deeper understanding of exactly how much the overlap of fractal surfaces at geological faults determines the observed earthquake statistics.

# Appendix A: The renormalization group approach

Chakrabarti and Stinchcombe [27], as we stated earlier, were the first to take up this Two Fractal Overlap model. They did a renormalization group calculation to obtain the frequency distribution of overlap magnitudes. To state the problem formally, Chakrabarti and Stinchcombe wanted to find the number density $n(\varepsilon)$ of earthquakes releasing energy

$\varepsilon$ in the Two Fractal Overlap model. To find this number density we need to find out the distribution $\rho(s)$ of the overlap magnitude $s$ between the two self-similar surfaces. We give here a short description of their method of solution. Let the sequence of generators that define our Cantor sets within the interval [0, 1] be denoted by $G_n$. This means: $G_0 =$ [0, 1], $G_1 \equiv RG_0 = [0, x] \cup [y, 1]$ ,..., $G_{n+1} = RG_n$,.... Of course in our work we have considered $x = 1/3$ and $y = 2/3$. If we represent the mass density of the set $G_n$ by $D_n(r)$, then $D_n(r) = 1$ if $r$ is in any of the occupied intervals and $D_n(r) = 0$ if $r$ is in any of the unoccupied intervals of the Cantor set. The required overlap between the sets at any generation $n$ is then given by the convolution integral:

$$s_n(r) = \int dr' D_n(r') D_n(r - r') \tag{A1}$$

This form applies to symmetric fractals. The generalized form of the integral would have the argument of the second $D_n$ in the integrand as $(r + r')$ of course. But symmetry implies $D_n(r) = D_n(-r)$. Some aspects of the convolution integral of two Cantor sets has been discussed in [53] in the context of band-width and band number transitions in quasicrystal models. This problem is however a more complex one. The method used by Chakrabarti and Stinchcombe is a generalization of the recursive scaling method used in [60] and gives a very direct solution to the problem. To understand this we need to have a look at Fig. A.1. One can express the overlap integral $s_1$ for the first generation by the projection of the shaded regions along the vertical diagonal in Fig. A.1 (A). The projections are of the type shown in Fig. A.1 (B). For $x = y \leq \frac{1}{3}$, the non-vanishing $s_1(r)$ regions do not overlap, and are symmetric on both sides with the slope of the middle curve being exactly double the slope of those on the sides. One can then easily check that the distribution $\rho_1(s)$ of overlap $s$ at this generation is given by Fig. A.1 (C), with both $w$ and $u$ greater than unity, maintaining the normalization of the probability $\rho_1$ with $wu = 5/3$.

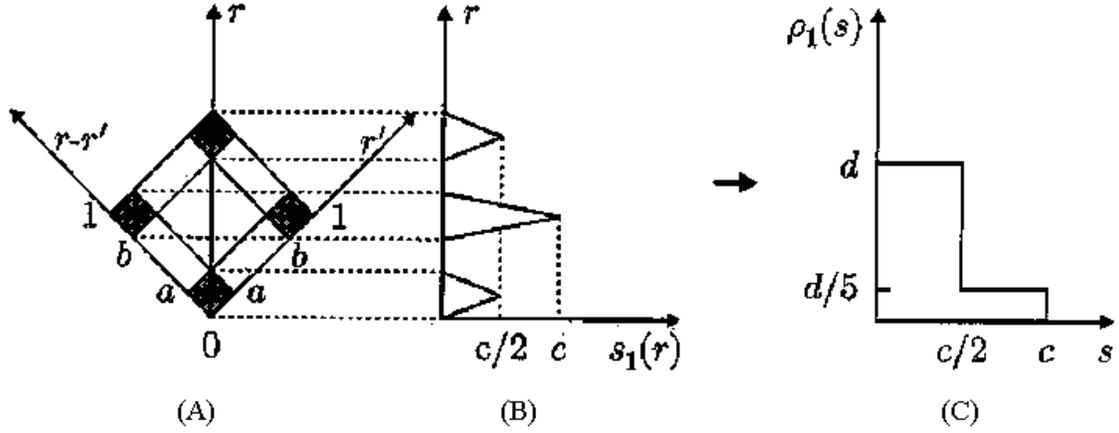

*Fig. A.1 (A) Two Cantor sets (of first generation) along the axes r' and r – r'. (B) This gives the overlap $s_1(r)$ along the diagonal. (C) The corresponding density $\rho_1(s)$ of the overlap s at this generation.*

For successive generations *n* the density $\rho_n(s)$ may be hence represented by a recursion relationship:

$$\rho_{n+1}(s) = R\rho_n(s) \equiv \frac{u}{5}\rho_n\left(\frac{s}{w}\right) + \frac{4u}{5}\rho_n\left(\frac{2s}{w}\right). \qquad (A2)$$

In the limit $n\to\infty$ the renormalization group (RG) equation, we will have the fixed point distribution $\rho^*(s)$ which will satisfy the recursion relationship as $\rho^*(s) = \tilde{R}\,\rho^*(s)$. If we assume $\rho^*(s) = s^{-\gamma}\,w\,\tilde{\rho}(s)$, from (A2) we will obtain $\left(\frac{u}{5}\right)w^\gamma + \left(\frac{4u}{5}\right)\left(\frac{w}{2}\right)^\gamma = 1$. Here $\tilde{\rho}(s)$ is an arbitrary modular function which also includes a logarithmic correction for large *s*. This agrees with the normalization condition $wu=5/3$ mentioned before for the choice $\gamma = 1$. This fixed point overlap frequency distribution is then given by:

$$\rho^*(s) \equiv \rho(s) \sim s^{-\gamma};\, \gamma = 1. \qquad (A3)$$

This can be checked by checking the behavior of an appropriately rescaled form of the actual distribution $\rho_n(s) = R^n\rho_0(s)$ for large values of *n*. This is the general result for all cases that Chakrabarti and Stinchcombe investigated and solved by the functional rescaling technique (with the log *s* correction for large *s,* renormalizing the total integrated distribution). The cases they investigated include non-random Cantor sets and the Sierpinski carpets (for slides along various directions). The power law statistics holds for the fixed point overlap frequency distribution function in all these cases. Fig. A.2 shows the $\rho_n(s)$ versus $s$ plots for progressively increasing generation number ($n$) values.

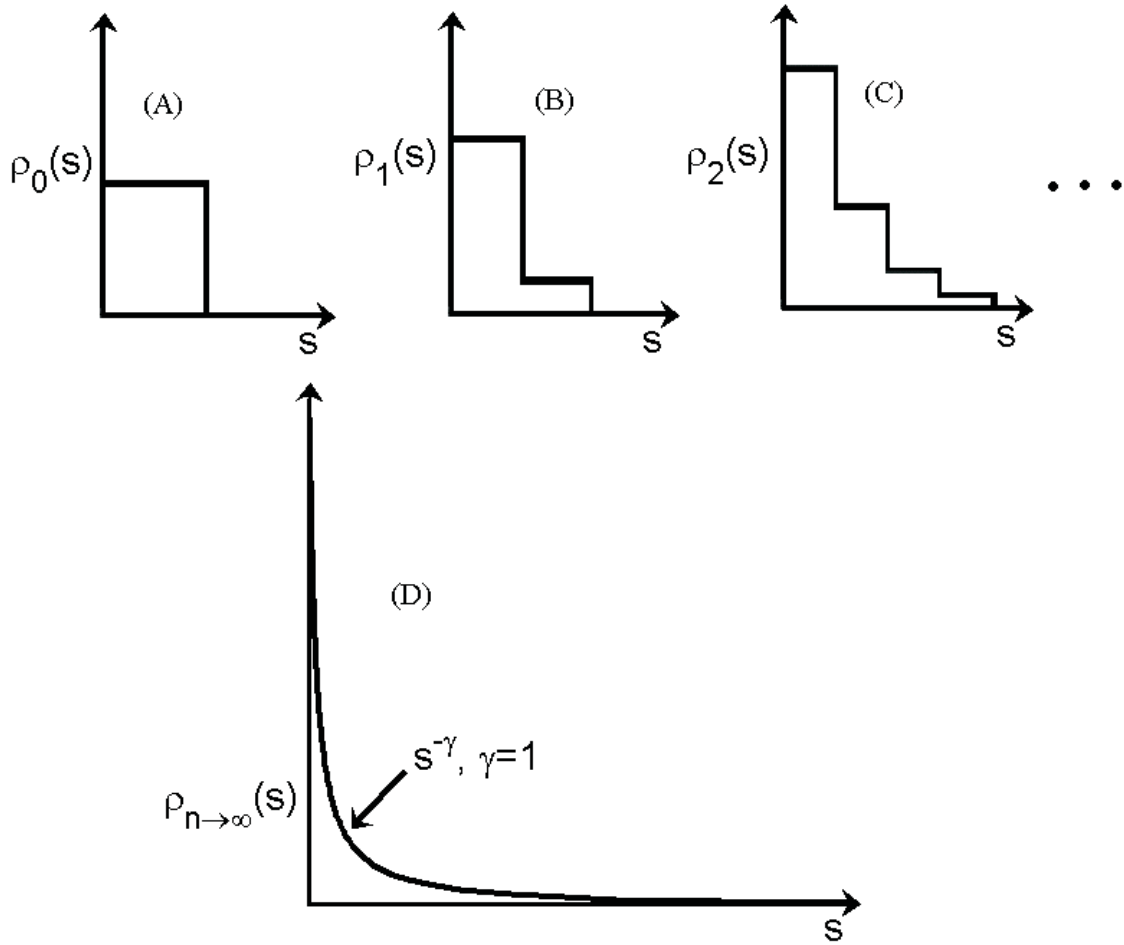

*Fig. A.2 The overlap densities $\rho_n(s)$ at various generations of the Cantor sets: at the zeroth (A), first (B), second (C) and at the infinite (or fixed point) (D) generations.*

# Appendix B: Details of the analysis of the Two Fractal Overlap Model time series

We present here the complete analysis of the Chakrabarti Stinchcombe model by Bhattacharya [28]. Periodic boundary conditions are employed to formulate the time series. The overlap magnitude is evaluated in terms of the number of pairs of non-empty intervals overlapping at a time. The overlap magnitude $O_n(t)$ can only assume values in a geometric progression given by $O_n(t) = 2^{n-k}$, $k = 0, 1, \ldots, n$. At $t=0$, all the occupied intervals of the Cantor sets overlap and hence clearly $O_n(t = 0) = 2^n$ and, due to the periodic boundary conditions, taking unit time to be the time required to take one step of size $1/3^n$ we obtain

$$O_n(t) = O_n(3^n - t), \quad 0 \leq t \leq 3^n \tag{B1}$$

due to the self similar structure of the finite generation Cantor set.

A detailed analysis of the time series reveals a straightforward recursive structure. The basic structure is derived from the following observation: If we simulate the overlap time series for the $n$-th generation, after $3^{n-1}$ time steps we have the overlap time series for the $(n-1)$-th generation. Again after $3^{n-2}$ time steps beginning from the $3^{n-1}$ time steps previously taken we have the overlap time series for the $(n-2)$-th generation and recursively so on (see Fig. 2.1). This is the self-affine structure built in the time series due to the structure of the Cantor set. This scheme yields in the rule

$$O_n\left(t = \sum_{r=1}^{k} 3^{n-r}\right) = 2^{n-k}; \quad k = 1,\ldots,n. \tag{B2}$$

There is however a finer recursive structure in the time series that leads to the analytical evolution of the number density distribution. At any given generation $n$, a pair of nearest line segments form a doublet and there are $2^{n-1}$ such doublets in the Cantor set. Within a given doublet, each segment is two time steps away from the other segment. This means that an overlap of $2^{n-1}$ occurs when one of the sets is moved two time steps relative to the other. Similarly, an overlap of magnitude $2^{n-1}$ also occurs if one considers a quartet and a relative shift of $2\times 3$ time steps between the two Cantor sets. Again we can consider an octet and a relative shift of $2\times 3^2$ time steps to obtain an overlap of magnitude $2^{n-1}$. In general if we consider pairs of blocks of $2^{r_1}$ nearest segments ($r_1 \leq n-1$), an overlap magnitude of $2^{n-1}$ occurs for a relative time shift of $2\times 3^{r_1}$ time steps:

$$O_n(t = 2\times 3^{r_1}) = 2^{n-1}; \quad r_1 = 0,\ldots,n-1. \tag{B3}$$

The complementary sequence is obtained using (B1). Now the next overlap magnitude is $2^{n-2}$. For each time step $t$ for which an overlap of $2^{n-1}$ segments occur, there are in general two subsequences of overlaps of $2^{n-2}$ segments that are mutually symmetric with respect to *t*; one preceding and the other succeeding *t*. Therefore the sequence of *t* values for which an overlap of $2^{n-2}$ segments occurs is determined by the sum of two terms; one from each of the two geometric progressions; one nested within the other:

$$O_n\left(2[3^{r_1} \pm 3^{r_2}]\right) = 2^{n-2}; \tag{B4}$$
$$r_1 = 1,......, n-1;$$
$$r_2 = 1,......, r_1 - 1.$$

Again the complementary sequence is given by (B1). In general, by induction, it is straight forward to deduce the sequence of time step values for which an overlap of $2^{n-k}$ segments occur ($1 \leq k \leq n$) within a period of $3^n$ steps. It is given by a sum of *k*-terms, one from each of *k* geometric series nested in succession:

$$O_n\left(2[3^{r_1} \pm 3^{r_2} \pm .... \pm 3^{r_{k-1}} \pm 3^{r_k}]\right) = 2^{n-k}; \tag{B5}$$
$$k = 1,......, n;$$
$$r_1 = k - 1,......, n - 1;$$
$$r_2 = k - 2,......, n - 2;$$
$$.\quad\quad.$$
$$.\quad\quad.$$
$$.\quad\quad.$$
$$r_{k-1} = 1,......, r_{k-2} - 1;$$
$$r_k = 1,......, r_{k-1} - 1;$$

For each value of *k* in the above sequence there is a complementary sequence due to the symmetry property stated by (B1). Thus the value of the overlap at zero time along with the symmetry property (B1) and (B5) determine our complete time series.

Of some considerable interest is the special case of unit overlaps. Unit overlaps occur when only one occupied interval of the Cantor sets are in a state of overlap. A unit overlap occurs when we put $n = k$ in (B5). The sequence of *t* values for which this happens is given by:

$$O_n\left(2[3^{n-1} \pm 3^{n-2} \pm....\pm 3^1 \pm 3^0]\right) = 2^{n-n} = 1. \tag{B6}$$

This gives $2^{n-1}$ occurrences of the unit overlap. An equal number of unit overlaps also occur due to the symmetry property (B1) in the complementary sequence arising due to the periodic boundary condition. So in all, for the *n*-th generation we have $2^n$ occurrences of the unit overlap. From (B1), (B3) and (B4) and the fact that $O_n(0) = 2^n = \max(O_n)$ we can easily deduce that:

$$N(2^n) = 1 \tag{B7}$$

$$N(2^{n-1}) = 2n \tag{B8}$$

$$N(2^{n-2}) = 2\sum_{r_1=1}^{n-1} 2r_1 = 2n(n-1) \tag{B9}$$

$$N(2^{n-3}) = 2\sum_{r_1}^{n-1} 2\sum_{r_2}^{r_1-1} 2r_2 = \frac{4}{3}n(n-1)(n-2) \tag{B10}$$

$$N(2^{n-4}) = 2\sum_{r_1}^{n-1} 2\sum_{r_2}^{r_1-1} 2\sum_{r_3}^{r_2-1} 2r_3 = \frac{2}{3}n(n-1)(n-2)(n-3) \tag{B11}$$

and so on where $N(O_n)$ denotes the number of times that an overlap of magnitude $O_n$ occurs within one period. From induction we arrive at the general formula

$$N(2^{n-k}) = 2\sum_{r_1=k-1}^{n-1} 2\sum_{r_2=k-2}^{r_1-1} \ldots 2\sum_{r_{k-1}=1}^{r_{k-2}-1} 2r_{k-1} = C_k n(n-1)(n-2)\ldots(n-k+1) = C_k \frac{n!}{(n-k)!} \tag{B12}$$

where $C_k$ is a constant. $C_k$ can be determined from the case of unit overlaps. If we put $n = k$ and keep $k$ constant in (B12) then we have the frequency of unit overlaps for the $k$-th generation. This comes out to be

$$N(2^{k-k}) = C_k k! \tag{B13}$$

On the other hand, from (B6) we had already derived that for the $n$-th generation the number of unit overlaps was $2^n$. So for the $k$-th we will have $2^k$ unit overlaps. This gives us the following:

$$C_k k! = 2^k. \tag{B14}$$

From (B14) $C_k$ comes out to be $\frac{2^k}{k!}$. Normalizing with $3^n$ we obtain the probability distribution function for overlap magnitudes as;

$$Pr(2^{n-k}) = \binom{n}{n-k}\left(\frac{1}{3}\right)^{n-k}\left(\frac{2}{3}\right)^k \tag{B15}$$

where $Pr(O_n) = \frac{N(O_n)}{3^n}$. Now, remembering that the overlap magnitude $2^{n-k}$ is proportional to energy we can put $\log_2 O_n = n - k = m$ where $m$ is the magnitude analog for the model. Then the frequency distribution for the model in terms of magnitude becomes

$$Pr(m) = \binom{n}{m}\left(\frac{1}{3}\right)^m\left(\frac{2}{3}\right)^{n-m} \tag{B16}$$

In the limit of large $n$ the Cantor set becomes a true mathematical fractal and we can have the standard normal approximation of (B16). Basically in the limit of large $n$ we can approximate $n!$ as

$$n! \cong \sqrt{2\pi n}\, n^n \exp(-n) \tag{B17}$$

which is the standard Sterling approximation [61]. Under the Sterling approximation the binomial distribution (B16) can be approximated as a normal distribution, given the mean $\mu$ and the standard deviation $\sigma$ of the binomial distribution, in the form:

$$F(h) = \frac{\exp(-h^2/2)}{\sqrt{2\pi}\sigma} \tag{B18}$$

where

$$h = \frac{m - \mu}{\sigma}. \tag{B19}$$

Of course $m$ is the variable representing magnitude that is the variable which is binomially distributed. The mean of the distribution in (B16) is $n/3$ and the standard deviation is $\sqrt{2n/9}$. Therefore we can write down (B16) as

$$F(m) = \frac{3}{2\sqrt{n\pi}} \exp\left[-\frac{9}{4}\frac{(m-n/3)^2}{n}\right]. \tag{B20}$$

As indicated in the text, to obtain the GR law analog from this distribution we have to integrate $F(m)$ from $m$ to $\infty$. Neglecting terms with coefficients of the order of $1/n\sqrt{n}$ and higher we obtain the cumulative distribution function for magnitude $m$ and above as

$$F_{cum}(m) = \frac{3}{2\sqrt{n\pi}} \exp(-n/4)\exp\left[-\frac{9m^2}{4n} + \frac{3m}{2}\right](m - n/3). \tag{B21}$$

Now, in the large magnitude limit, as the magnitude $m$ in the model cannot exceed $n$, the term $m^2/n \sim m$ and hence effectively (B21) becomes

$$F_{cum}(m) = \frac{3}{2\sqrt{n\pi}} \exp(-n/4) \exp\left[-\frac{3m}{4}\right](m - n/3). \tag{B22}$$

On taking log of both sides of (B22) we obtain

$$\log F_{cum}(m) = A - \frac{3}{4}m + \log(m - n/3) \tag{B23}$$

where $A$ is a constant depending on $n$. This is the GR law analog for the model which clearly holds for the high magnitude end of the distribution. This derivation of the GR like cumulative frequency distribution is a new analytical extension of the work presented in [28]. The important realization that led to this development was that the GR law was in reality a cumulative statistics and it is log-linear relationship in cumulative number versus magnitude rather than a log-linear relationship between number density and energy released.

# Appendix C: Overlap magnitude distributions for different fractals

Here we present the overlap magnitude distributions of some fractals other than the regular Cantor set described in the main text. We review the numerical analysis undertaken in [62] where the contact area distributions between two fractal surfaces have been studied for various types of fractals of different fractal dimensions. The variations in overlap magnitude ($O_n$) were studied for two self-similar fractals, both of the same fractal dimension ($d_f$) and the same generation $n$, one sliding with uniform relative velocity over the other (which is really the Two Fractal Overlap model for fractals other than the regular Cantor set). The main objective of [62] was to formulate the probability distribution $Pr(O_n)$ of the overlap magnitude $O_n$. Below we present a brief discussion of the overlap magnitude distributions obtained in [62] with several different fractals, namely: 1) random Cantor sets 2) regular and random Sierpinski gaskets on a square lattice and 3) percolating clusters embedded in two dimensions. A universal scaling behaviour of the distribution was found in [62] of the form:

$$P'(O_n') = L^\alpha \Pr(O_n, L); O_n' = O_n L^{-\alpha}, \tag{C1}$$

where $L$ denotes the finite size of the fractal and the exponent $\alpha = 2(d_f - d)$. Here $d_f$ is the fractal or mass dimension of the fractal and $d$ is the embedding dimension of the fractal. Also the overlap distribution function $Pr(O_n)$, and hence also the scaled distribution $P'(O_n')$, is seen to decay with $O_n$ or $O_n'$ following a power law with the exponent value equal to the embedding dimension $d$ for both regular and random Cantor sets and gaskets:

$$Pr(O_n) \sim O_n^{-\beta}; \beta = d. \tag{C2}$$

For the percolating clusters, however, the overlap distribution takes a Gaussian form. The normalization restriction on both $Pr(O_n)$ and $O_n$ ensures the same scaling exponent $\alpha$ for both. The result for the regular Cantor sets has already been discussed at length in Appendix A and another approach may also be found in [28]. Hence we do not discuss regular Cantor sets anymore and take up the cases of the other fractals considered in [62].

A) Random Cantor sets

Two types of random Cantor sets were considered for formulation of the distribution function $Pr(O_n)$ in [62]. Random Cantor sets of dimension $d_f = \ln 2/\ln 3$ and random Cantor sets of dimension $d_f = \ln 4/\ln 5$. A random Cantor set of dimension $d_f = \ln 2/\ln 3$ can be created by first dividing the generator line segment, which is usually of unit length at generation zero ($n = 0$), into three equal parts and then removing any of the one-third portions randomly at every generation (see Fig. C.1). Here the structure of the sets change with configurations as randomness is involved. The overlap between any two such sets at finite generation $n$ having same dimension but of different configurations are considered. Clearly the distribution of overlap sizes $O_n$ changes with each pair of configurations and hence the distribution $Pr(O_n)$ determined is the average distribution. The finite size $L$ of the Cantor set with $d_f = \ln 2/\ln 3$ is $L = 3^n$ at generation $n$ and for $d_f = \ln 4/\ln 5$ we have $L = 5^n$. Overlap distributions $Pr(O_n, L)$ are fitted to the scaling forms (C1) and (C2). The average distributions $Pr(O_n, L)$ were determined using 500 such configurations for each of the two Cantor sets. The relevant plots are shown in Fig. C.2. The plots in Fig. C.2 indicate the validity of (C1) and (C2) in the limit of large $n$ (or large $L$).

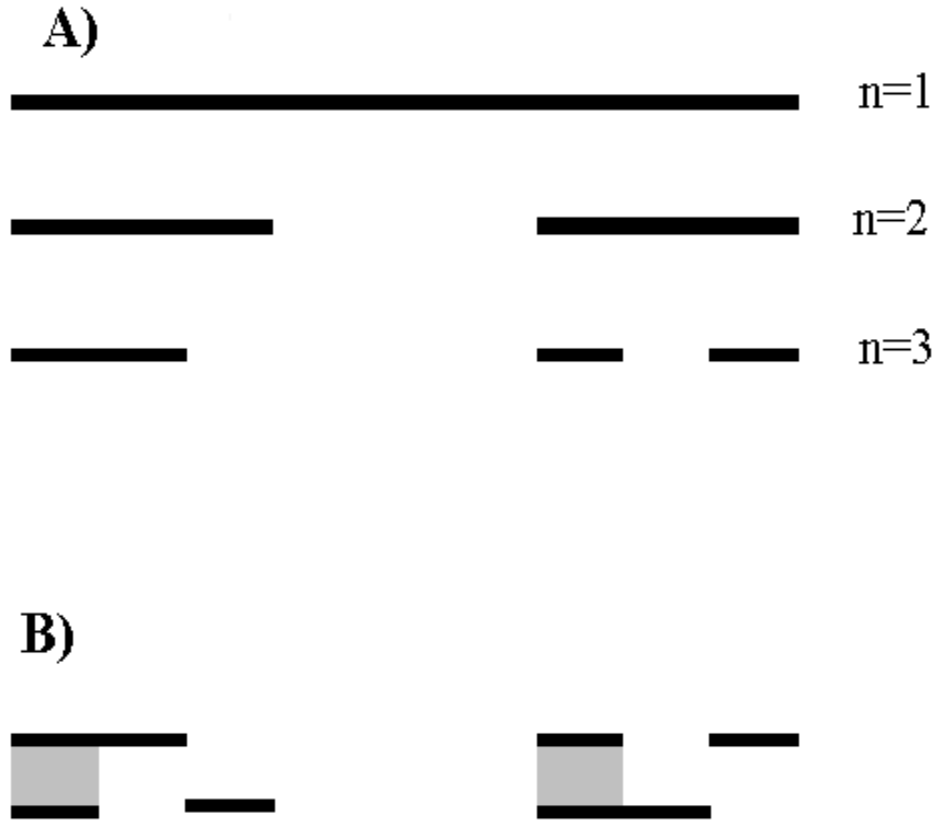

*Fig. C.1 A) A random Cantor set of dimension ln 2/ln 3 in the first three generations. B) Overlap of two such random Cantor sets of the same generation (here n=3). Shaded regions are the regions of overlap at the given time step (here t=0) (adapted from [62]). As in the text the number of such overlapping non-empty intervals gives the measure of the overlap magnitude. Therefore the overlap magnitude is the number of overlapping non-empty intervals of the two Cantor sets sliding over each other.*

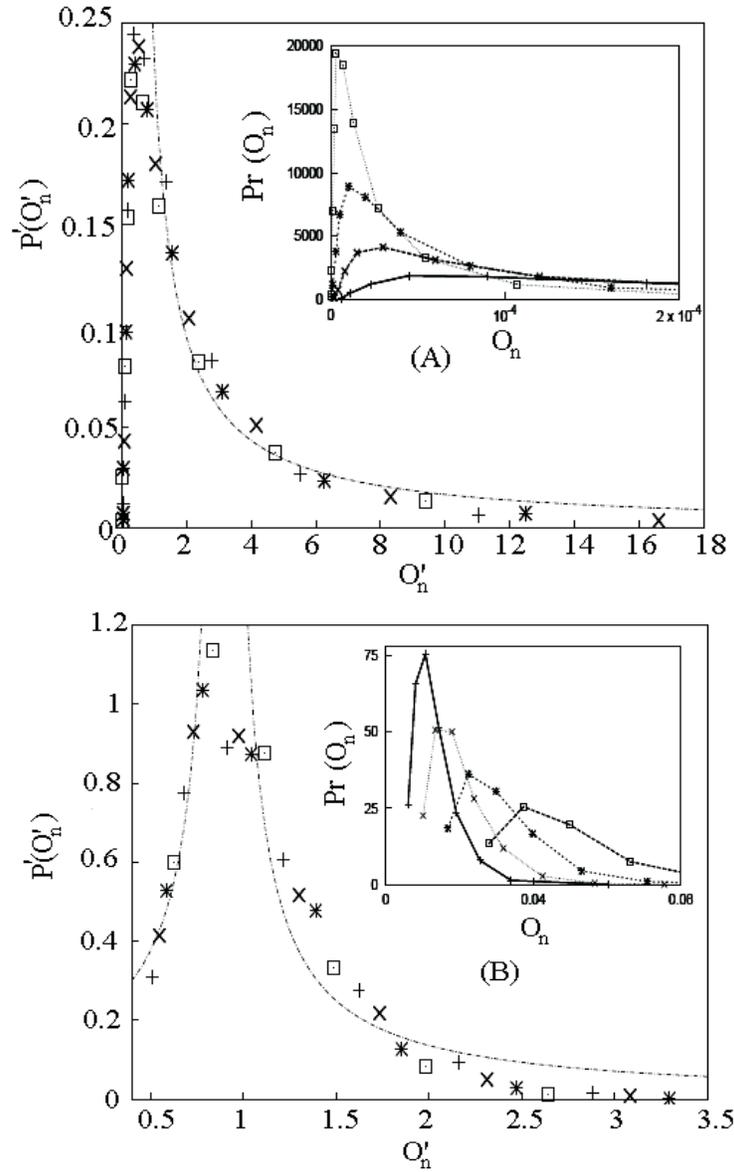

*Fig. C.2 Plots of the scaled distribution (configurationally averaged) $P'(O_n') = L^\alpha Pr(O_n, L)$ versus scaled overlap magnitude $O_n' = O_n L^{-\alpha}$ and in the inset the unscaled distribution for: A) random Cantor set with $d_f = \ln 2/\ln 3$ at various finite generations: n = 11 (plus), n = 12 (cross), n = 13 (star) and n = 14 (square); B) random Cantor set with $d_f = \ln 4/\ln 5$ at various finite generations: n = 7 (square), n = 8 (star), n = 9 (cross) and n = 10 (plus). The exponent $\alpha = 2(d_f - d)$ given embedding dimension d and fractal dimension $d_f$. The dotted lines indicate the best fit curves of the form $\eta(x - \delta)^{-d}$ where d = 1 (adapted from [62]).*

B) Regular Sierpinski gaskets

In the case of regular fractals we once again use periodic boundary conditions similar to the one employed in the Two Fractal Overlap model to avoid end effects. Sierpinski gaskets have fractal dimension ln 3/ln 2 (see Fig. C.3). Finite generations were considered. Since no randomness is involved no configurational averaging is required. Once again the overlap distribution $Pr(O_n, L)$ is fitted to the scaling forms (C1) and (C2). The results are shown in Fig. C.4.

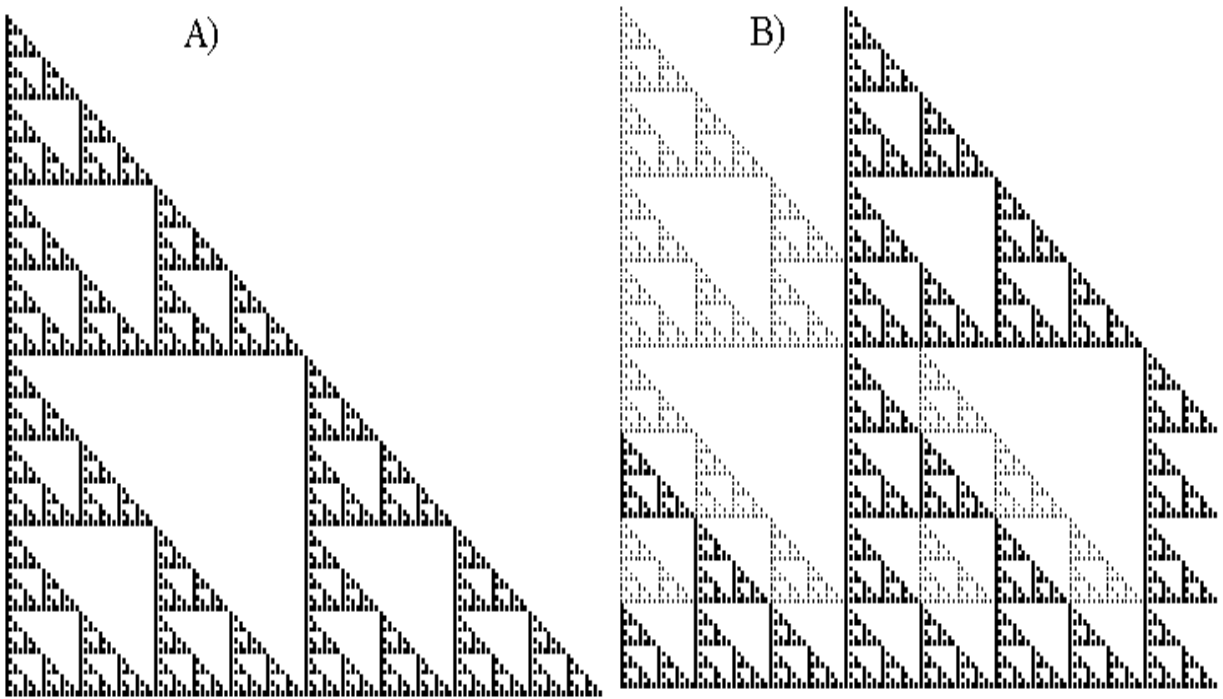

*Fig. C.3 A) A regular gasket of dimension $d_f$ = ln 3/ln 2 at generation n = 7. B) Overlap of two regular gaskets, both at the same generation n = 7, is shown as one translates over the other. Periodic boundary condition on the translated gasket (adapted from* [62]*).*

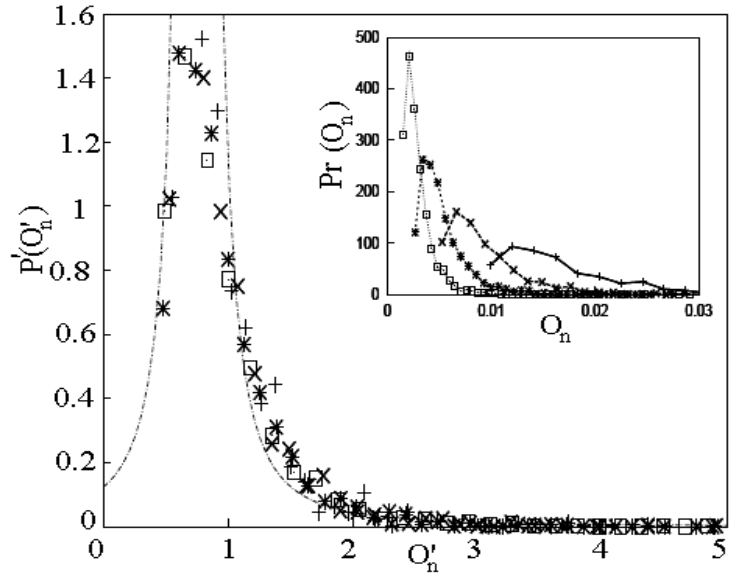

*Fig. C.4 Plots of the scaled distribution $P'(O_n') = L^{\alpha} \Pr(O_n, L)$ versus scaled overlap magnitude $O_n' = O_n L^{-\alpha}$ and in the inset the unscaled distribution for the regular gasket with $d_f = \ln 3/\ln 2$ at various finite generations: n = 7 (plus), n = 8 (cross), n = 9 (star) and n = 10 (square). The exponent $\alpha = 2(d_f - d)$ given embedding dimension d and fractal dimension $d_f$. The dotted lines indicate the best fit curves of the form $\eta (x - \delta)^{-d}$ where d = 2 (adapted from [62]).*

## C) Random Sierpinski gaskets

For random gaskets exactly the same methodology was used as for the random Cantor sets. Different configurations are taken and the configurationally averaged distribution is determined using 500 different configurations of the pair of upper and lower gaskets. Two such random gaskets and their overlaps are shown in Fig. C.5. The plots for the overlap distributions for various generations (configurationally averaged) are shown in

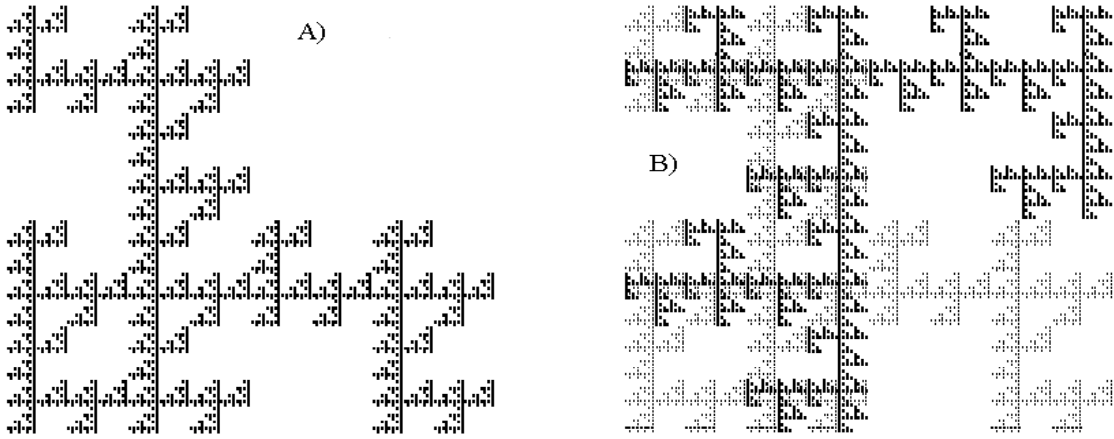

*Fig. C.5 A) A random realization of a gasket of dimension $d_f = \ln 3/\ln 2$ at generation n = 7. B) The overlap of two random gaskets of the same dimension and generation (n = 7) but of different configurations (adapted from [62]).*

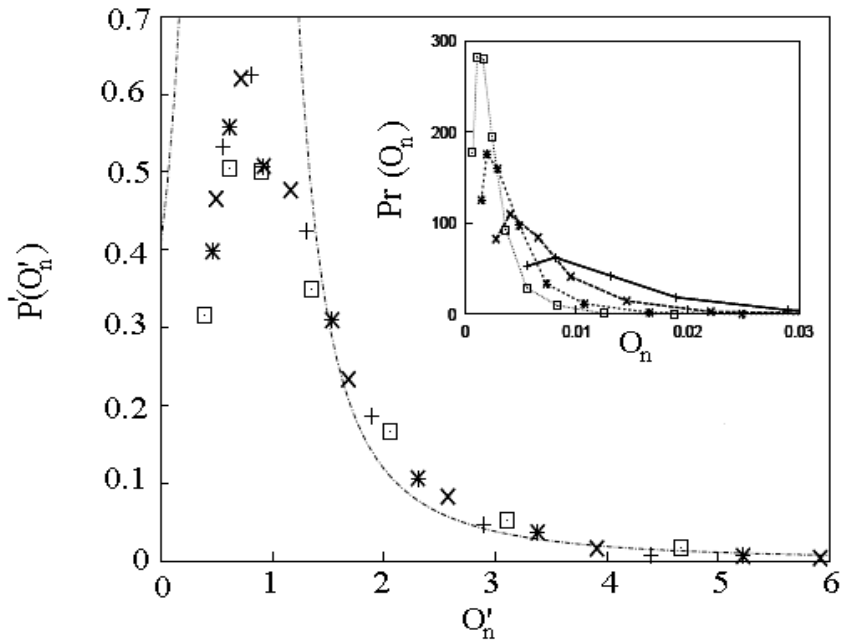

*Fig. C.6 Plots of the scaled distribution (configurationally averaged) $P'(O_n') = L^\alpha \Pr(O_n, L)$ versus scaled overlap magnitude $O_n' = O_n L^{-\alpha}$ and in the inset the unscaled distribution for random gaskets with $d_f = \ln 3/\ln 2$ at various finite generations: n = 8 (plus), n = 9 (cross), n = 10 (star) and n = 11 (square). The dotted lines indicate the best fit curves of the form $\eta (x - \delta)^{-d}$ where d = 2 (adapted from [62]).*

Fig. C.6. Again of course $Pr(O_n, L)$ is configurationally averaged and fitted to the scaling forms (C1) and (C2). Fitted curves are also shown in Fig. C.6.

D) Percolating Clusters in a square lattice

Percolating clusters are very well characterized random fractals (for a detailed discussion see [63]). Efficient and widely used algorithms are available to generate such fractals. Although many detailed features of the clusters change with the changes in the parent fractals the subtle features of the overlap magnitude distribution function remains unchanged. Several site percolating clusters were numerically generated at the percolation threshold ($p_c$ = 0.5927 [62]) on square lattices of linear size $L$ by using the Hoshen-Kopelman algorithm [63, 64]. To enumerate the overlap set for percolating clusters the number of sites $N$ which are occupied in both the clusters (see Fig. C.7) are counted. Then the overlap size $O_n$ is given as $O_n = N/L^d$ where $d$ is again the embedding dimension (which is two here). Of course $n$ here denotes the linear number of lattice vertices or sites available, that is $n = L$ and there is a total number of $L^2$ sites. As the realizations of the fractal were varied keeping the fractal dimension $d_f$ the same $O_n$ changed and the distribution $Pr(O_n, L)$ was determined. It was seen that the distribution shifts continuously as $L$ increases and has a finite width which diminishes very slowly with $L$. This shows that the emerging length scale associated with $O_n$ is no more a constant but it depends on $L$. This is due to the fractal nature of the original clusters where the occupations of sites are no longer truly random occurrences but are correlated [62]. Hence for a system of size $L$ the probability of occupation grows as $L^{d_f}$ for any of the fractals and as $L^{2(d_f-d)}$ for the overlap set. If this is the reason for the $L$ dependence of the width and shift in $Pr(O_n, L)$ then the distributions for different $L$ values will follow the scaling law in (C1).

From the overlap between all the pairs of cluster configurations (typically around 500 for $L$ = 400) the distribution $Pr(O_n, L)$ was determined. The data were binned to facilitate storage and to smoothen the distribution (Fig. C.8 (A)). The nature of the distribution $Pr(O_n)$ was also examined for percolation clusters generated at values of the lattice occupation probability $p$ higher than the critical value $p_c$ for the square lattice.

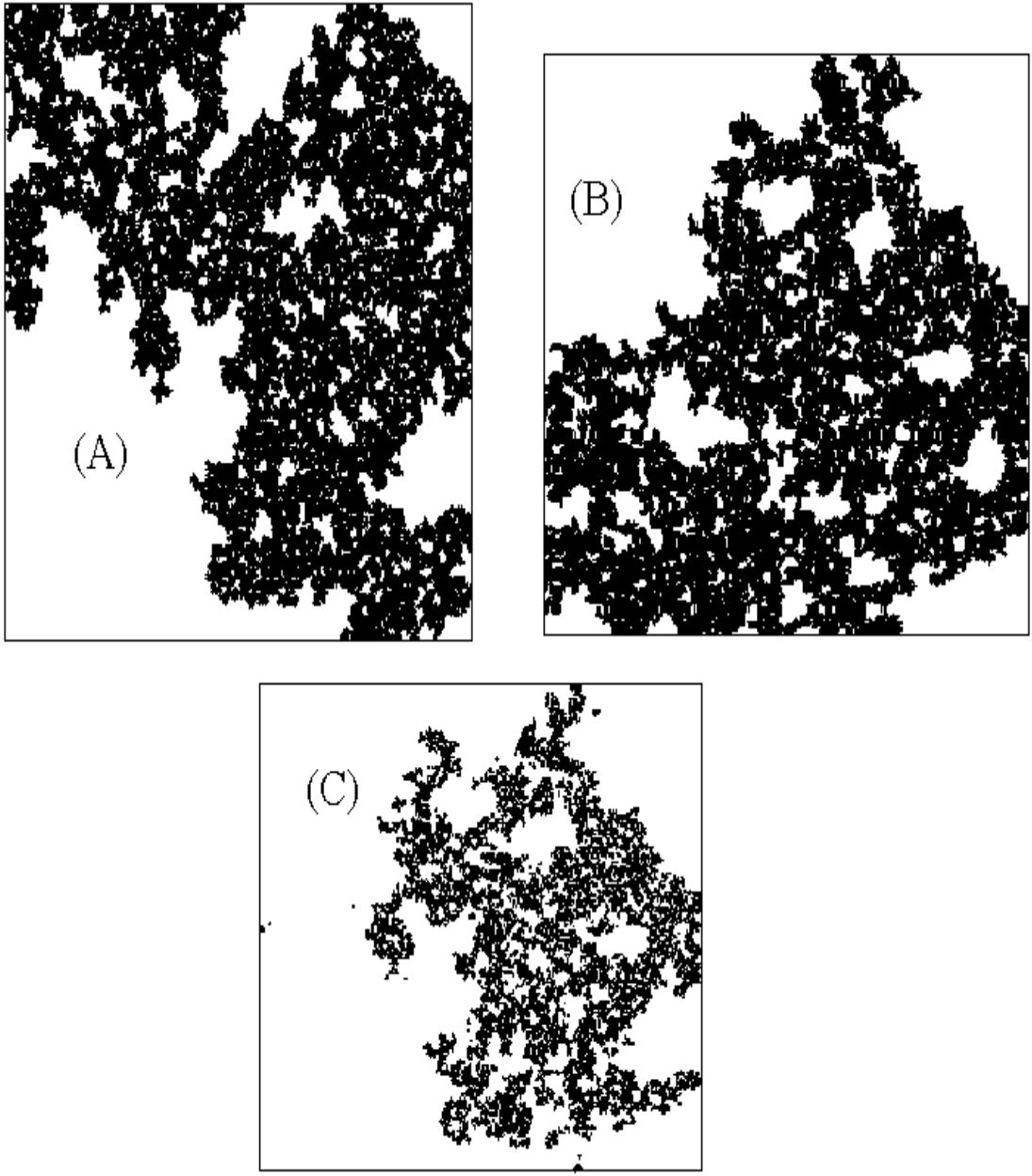

*Fig. C.7 The overlap between two percolation clusters. A) and B) are two realizations of the same percolating fractal on a square lattice ($d_f = 1.89$). C) Their overlap set. It is interesting to note that the overlap set need not be connected (adapted from [62]).*

The resultant distributions become delta functions and the height of the peaks increase with the increase in system size $L$ (Fig. C.8 (B)).

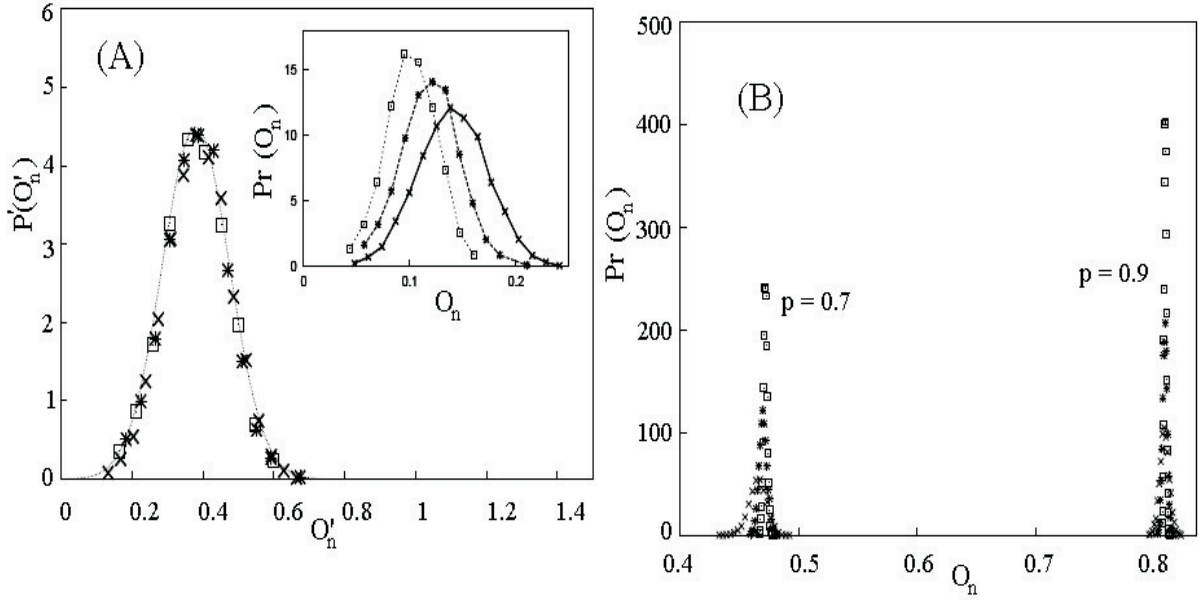

*Fig. C.8 A) Plot of the scaled distribution $P'(O_n') = L^\alpha Pr(O_n, L)$ versus scaled overlap magnitude $O_n' = O_n L^{-\alpha}$ for percolating clusters grown with probability $p=p_c=0.5927$ on a square lattice ($d_f = 1.89$) of finite linear extent: $L = 100$ (cross), $L = 200$ (star), $L = 400$ (square). The exponent $\alpha = 2(d_f - d)$ given embedding dimension $d$ and fractal dimension $d_f$. The dotted line indicates the best fit curves of the form $\eta \cdot \exp(-(x - \delta)^{-2.0}/\zeta)$; where $\gamma$, $\delta$ and $\zeta$ are constants. Here $\eta = 0.45$, $\delta = 0.38$ and $\zeta = 0.018$. Inset shows the unscaled distribution $Pr(O_n)$ plotted versus unscaled overlap magnitude $O_n$. B) The unscaled distribution $Pr(O_n)$ versus unscaled overlap magnitude $O_n$ for percolation clusters grown with probability of occupation of site $p_{site}$ greater than $p_c$ ($p_{site} > p_c$, $p_{site} = 0.7$, $p_{site} = 0.9$) on a square lattice ($d_f = 1.89$) of finite linear extent: $L = 100$ (cross), $L = 200$ (star), $L = 400$ (square). The distributions are clearly delta functions and the peak amplitude increases with linear lattice size L (adapted from [62]).*